\documentclass[aps,prfluids,showkeys,superscriptaddress]{revtex4-2}

\usepackage[utf8]{inputenc} 
\usepackage[T1]{fontenc}    
\usepackage{url}           
\usepackage{booktabs}       
\usepackage{amsfonts}       
\usepackage{nicefrac}       
\usepackage{microtype}      
\usepackage{float}
\usepackage{caption}
\usepackage{caption}
\usepackage{subcaption}
\usepackage{amssymb,amsmath,amsthm}
\usepackage{changepage}
\usepackage{color,graphicx,float}
\usepackage{bm,enumerate}
\usepackage{alltt,listings,tikz}
\usepackage[all]{xy}
\usepackage{hyperref}
\usepackage{gensymb}
\usepgflibrary{shadings}

\newcommand{\bE}{\mathbf{E}}

\newcommand{\bT}{\mathbf{T}}
\newcommand{\bI}{\mathbf{I}}

\newcommand{\hx}{\hat{x}}
\newcommand{\hy}{\hat{y}}
\newcommand{\hu}{\hat{u}}
\newcommand{\hv}{\hat{v}}
\newcommand{\hh}{\hat{h}}

\newcommand{\ie}{\emph{i.e.}}
\newcommand{\eg}{\emph{e.g.}}

\begin{document}

\title{An electro-hydrodynamics modeling of droplet actuation on solid surface by surfactant-mediated electro-dewetting}
\author{Weiqi Chu}
\affiliation{Department of Mathematics, University of California Los Angeles}

\author{Hangjie Ji}
\affiliation{Department of Mathematics, North Carolina State University}

\author{Qining Wang}
\author{Chang-Jin ``CJ'' Kim}
\affiliation{Mechanical and Aerospace Engineering Department, University of California Los Angeles}

\author{Andrea L. Bertozzi}
\affiliation{Department of Mathematics, University of California Los Angeles}
\affiliation{Mechanical and Aerospace Engineering Department, University of California Los Angeles}

\begin{abstract}
We propose an electro-hydrodynamics model to describe the dynamic evolution of a slender drop containing a dilute ionic surfactant on a naturally wettable surface, with a varying external electric field. This unified model reproduces fundamental microfluidic operations controlled by electrical signals, including dewetting, rewetting, and droplet shifting. In this paper, lubrication theory analysis and numerical simulations illustrate how to electrically control the wettability of surface via the charged surfactant.
Our numerical results show that electric field promotes dewetting by attracting ionic surfactants onto the transition thin-film region and promotes rewetting by attracting them away from the region.

\end{abstract}

\keywords{electro-dewetting, electro-wetting, microfluidics, droplet actuation, contact angle, ionic surfactant, lubrication theory}

\maketitle

\section{Introduction}\label{sec: intro}
In recent years, digital microfluidics (DMF) \cite{kim2001micropumping}, which allows manipulation of liquid droplets individually and independently \cite{li2020current}, has been intensively studied as an important liquid-handling technology \cite{choi2012digital} for lab-on-a-chip devices \cite{samiei2016review,abdelgawad2009digital} and many other applications \cite{chiu2012liquid,sen2008microscale,cheng2010active,nelson2011miniature,cha2016thermal}.
Among the different mechanisms to actuate droplets for DMF, electrowetting \cite{beni1981electro} in the form of electrowetting-on-dielectric (EWOD) \cite{kim2001micropumping} is the most widely used actuation mechanism due to its great flexibility and functionality \cite{quilliet2001electrowetting,nelson2012droplet,mugele2005electrowetting}. Electrowetting enhances the apparent wettability of liquid on a substrate with an electric potential, decreasing the contact angle and, as a result, deforming droplet shape on the substrate.
DMF is deemed feasible when basic droplet-manipulation operations, such as droplet generation, translation, splitting, and merging, are obtained with EWOD actuation \cite{cho2003creating}.

Pioneering work by Li et al. \cite{li2019ionic} verified that it is possible to diminish the wettability of liquid on a substrate using an electric potential, which is referred to as \emph{electro-dewetting} as opposite to electro-wetting. The trick is to add a dilute concentration of ionic surfactants to the liquid and electrically attract surfactants onto a hydrophilic substrate so that surfactants turn the hydrophilic surface into a hydrophobic one. 
The electro-dewetting is useful because it allows DMF, including directional movement, droplet splitting, etc., without requiring the dielectric and hydrophobic coating as well as the high voltages for EWOD. 
It is worth clarifying here that electro-dewetting (or electro-rewetting when an opposite voltage applied) \cite{li2019ionic} alters wettability by adding and removing surfactant molecules on the substrate and is intrinsically different from  electro-wetting \cite{nelson2012droplet}, which increases apparent wettability by attracting the liquid toward the substrate electrostatically. 
As a result, existing mathematical modeling studies of electro-wetting \cite{craster2005electrically,gao2021surfactant} are unsuitable for electro-dewetting or electro-rewetting.
It is intriguing to develop mathematical models to explain electro-dewetting and electro-rewetting in consideration of both liquid and surfactant dynamics with the presence of external electric force. 

Classical lubrication models have been extensively studied for the dynamics of thin liquid droplets spreading on a substrate \cite{craster2009dynamics,oron1997long}. Using a long-wave (or lubrication) approximation, these models typically describe the fluid flow driven by the interplay between Marangoni stresses \cite{dukler2020theory,BERTOZZI1999431}, surface tension \cite{myers}, evaporation \cite{ajaev2005evolution,ji2018instability}, and/or intermolecular forces \cite{hocking93}. Depositing surfactants over the surface of droplets can lead to a rich variety of morphological changes and pattern formation, including droplet spreading, fingering formation, and rupture instabilities.
In 2004, Warner et al. \cite{warner2004fingering,doi:10.1063/1.1763408} developed coupled lubrication equations for the liquid film thickness and the surfactant concentration for both cases of insoluble and soluble surfactants. To characterize droplet autophobing, Craster and Matar \cite{craster2007autophobing} also studied a model that incorporates the effect of surfactant on the wettability of a surface. More recently, a gradient dynamics model for liquid films with soluble surfactant was proposed by Thiele et al. \cite{thiele2016gradient}.

On the other hand, the effects of electric fields on the thin film interfacial instability, with or without surfactants, have also been investigated. In 2003, Warner et al. \cite{warner2003pattern} developed the lubrication approximation for the dynamic evolution of charged surfactant on thin liquid films. For an extensive overview of electro-hydrodynamics, we refer to \cite{saville1997electrohydrodynamics}. The electrically induced pattern formation in thin leaky dielectric films in the absence of surfactants was studied by Craster and Matar \cite{craster2005electrically}. Some recent works \cite{nganguia2019effects,poddar_mandal_bandopadhyay_chakraborty_2018,PhysRevFluids.6.064004} also focused on the effects of an electric field on surfactant-laden droplets without substrate support. With that being said, there still lacks a quantitative explanation of surfactant transport and contact angle evolution with the presence of electric field, which is the main venue to microfluidic operations in electro-dewetting.

In this paper, instead of interpreting the electric influence as a body force directly applied to the liquid, we propose an electro-hydrodynamic model which takes the dynamics of ionic surfactants under an electric field into consideration, and in turn, investigate the essential role of surfactants to alter surface tension and intermolecular forces \cite{cho2003creating, nelson2012droplet}. 
In this paper, we focus on establishing a unified mathematical model that describes the dynamics of thin fluid and insoluble surfactants in the presence of an electric field and compare our model with experimental results including contact angle changes and directional droplet movement. 

The structure of this paper is as follows. Section \ref{sec: math} proposes a fluid mechanics-based model describing the dynamics of a single droplet containing cationic surfactants with an open configuration. Based on the lubrication theory, we arrive at two coupled governing equations for the film thickness $h$ and the interface surfactant concentration $\Gamma$ that characterize the dynamics of the fluid system given an external electric field.
We also use a dimension reduction approach to compute the electric distribution efficiently.
In Section \ref{sec: experiments}, we describe the experimental setup and procedures to obtain contact angle data.
In Section \ref{sec: numerics}, we present the numerical results of droplet profiles when applying external voltages to the system over time. We compare the numerical simulations against experimental results and replicate droplet responses, including electro-dewetting, electro-rewetting, autophobing, and droplet shifting. We conclude our work in Section \ref{sec: conclusions}.

\section{Mathematical model}\label{sec: math}
\begin{figure}[htpb] 
\captionsetup{justification=centering}
\includegraphics[width=0.8\textwidth]{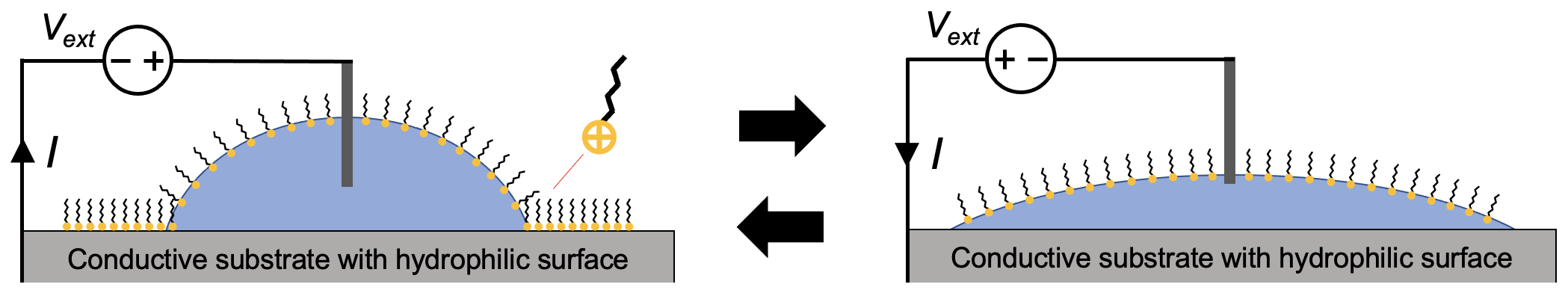}
\caption{A schematic diagram of electro-dewetting (left) and electro-rewetting (right) with sessile droplet configuration.}
\label{fig: dropletdiagram}
\end{figure}

Let us consider a sessile droplet containing positively charged (cationic) surfactants placed on a silicon wafer covered with a very thin (1-2 nm) layer of native silicon dioxide, which is hydrophilic, following \cite{li2019ionic}. After the droplet rests to its steady-state configuration, an electric field is turned on by means of top and bottom electrodes: the top electrode is a pin-shape electrode that can be inserted into the drop, and the bottom substrate consists of electrodes covered with the thin native oxide, as shown in Fig.~\ref{fig: dropletdiagram}.
In one electrode configuration, the substrate electrode imposes a time-dependent homogeneous potential, while the top pin electrode is grounded. When a negative voltage is applied on the substrate, the contact angle increases and leads the droplet to retreat. In contrast, when a positive voltage is applied to the substrate, the contact angle decreases and leads to droplet spreading \cite{li2019ionic}.
In another electrode configuration, multiple electrodes are placed on the substrate (as shown later in Fig.~\ref{fig: case2}) and applying a nonhomogeneous potential on the substrate leads to droplet shifting \cite{li2019ionic}.

\begin{figure}[htpb]
\textbf{ \quad macroscopic description \qquad \qquad \qquad mesoscopic description} \par\medskip
\includegraphics[width=0.56\textwidth]{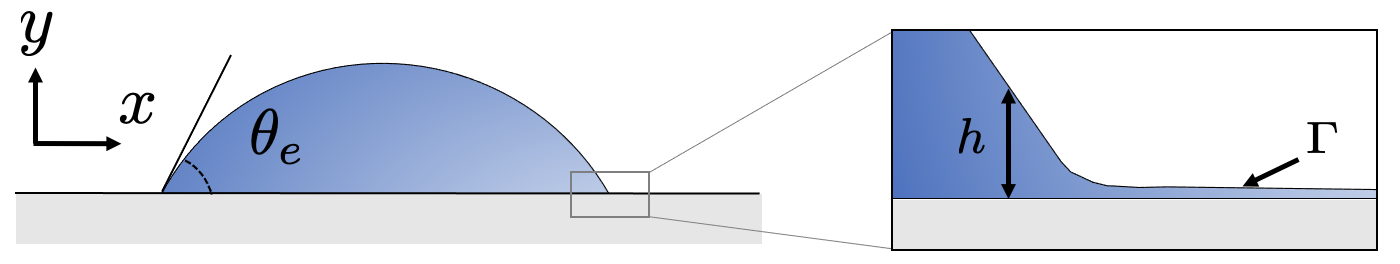}
\caption{Schematic diagrams of a droplet under macroscopic and mesoscopic descriptions. Left: $\theta_{e}$ is the equilibrium contact angle between the plane tangent to the droplet and a smooth substrate at the points of the three-phase (\ie, solid/liquid/gas). Right: a zoom-in picture near the triple point using a mesoscopic description, where film thickness $h$ and surfactant concentration $\Gamma$ are defined on the whole substrate surface.
}
\label{fig: mesodiagoram}
\end{figure}

The study on contact angles of sessile droplets on smooth surfaces dates back to the 19th century, where Young related the contact angle to surface tension by minimizing the free energy of the system from a macroscopic point of view \cite{young1805iii}. With the absence of external force, such as gravity or electric fields, a droplet spreads to its steady-state configuration (Fig.~\ref{fig: mesodiagoram} left) characterized by Young's equation,
\begin{equation}
\cos \theta_{e} = \frac{\sigma_{sg}-\sigma_{sl}}{\sigma_{lg}},
\end{equation}
where $\sigma$ represents the surface tension between solid (s), gas (g) and liquid (l).
In the mesoscopic picture, the substrate is covered by an equilibrium adsorption layer (Fig.~\ref{fig: mesodiagoram} right) and the contact angle is determined by a wetting energy dependent on the film thickness \cite{warner2002dewetting}. With the presence of surfactants, the consistency between macroscopic and mesoscopic descriptions is satisfied by relating the interfacial tension and wetting energy to the surfactant concentration \cite{thiele2018equilibrium}.
Motivated by the autophobing study \cite{craster2007autophobing} and an essential mesoscopic structure of the interphase boundary \cite{pismen2002mesoscopic}, we use the mesoscopic description and assume that the solid is always covered by a thin adsorption layer in our model. Furthermore, given the small volume (microliter scale) of the droplets studied, we also neglect the effect of gravity. We aim to develop a spreading theory that takes into account the Marangoni flow, electric forces, and intermolecular forces to describe various phenomena of droplet motions, including autophobing, dewetting, rewetting, and droplet shifting. The calculation of droplet shapes with the presence of gravity can be found in \cite{lubarda2011analysis}, which involves numerical integration of a second-order nonlinear differential equation.

\subsection{Governing equations}
Let $\bm{u} = (u,v)$ be the $x$- and $y$- components of the velocity field as shown in Fig.~\ref{fig: mesodiagoram}. We consider the equations for the flow dynamics of an incompressible, Newtonian drop, 
\begin{equation}
\begin{aligned}
\nabla \cdot \bm{u} &= 0\,, \\
\rho_m \left( \frac{\partial \bm{u}}{\partial t} + \bm{u} \cdot \nabla \bm{u}\right) &= -\nabla \left( p - \Pi \right) + \mu\nabla^{2}\bm{u}\,.
\end{aligned}
\label{eq: ns}
\end{equation}
where $\rho_m$ is the mass density, $\mu$ is the dynamic viscosity, $p$ is the pressure, and $\Pi$ is the disjoining pressure that serves as the supplementary pressure as a result of the molecular force interactions between the ultra-thin films and the solid surface \cite{bonn2009wetting}. 
The liquid-gas interface separates the whole geometric domain into two parts, $\Omega_L$ and $\Omega_G$, which represent the liquid and gas regions respectively. The film thickness $h(x,t)$ follows the continuity equation for incompressible fluids at the free surface $y = h(x,t)$,
\begin{equation}
	\frac{\partial h}{\partial t} + u \frac{\partial h}{\partial x} = v.
	\label{eq: freebd}
\end{equation}
We assume the surfactants are insoluble and only exist on the air-liquid interface and further denote $\Gamma(x,t)$ as the surfactant concentration. In reality, the electro-dewetting experiments use an electric field to manipulate the adsorption of ionic surfactant molecules on the solid surface, which in turn affects the wettability and steady-state contact angles \cite{li2019ionic}. 
In order to account for the influence of surfactants on the solid surfaces, we use a surfactant-dependent disjoining pressure $\Pi(h,\Gamma)$ \cite{beacham2009surfactant,craster2007autophobing}. The disjoining pressure is significant near the triple point region, in which we assume the water-air surfactants induce the same effect as the solid surfactants because the film thickness is small. 
We refer readers to the review paper \cite{oron1997long} for detailed discussions of disjoining pressure. Motivated by \cite{craster2007autophobing,schwartz1998simulation}, we choose a form for the disjoining pressure
\begin{align}
\Pi(h,\Gamma) &= \frac{A_H}{6\pi h_{\infty}^n} \left[ \left( \frac{h_{\infty}}{h} \right)^n - \left(1+\frac{\Gamma}{\Gamma_\infty}\right)^k\left( \frac{h_{\infty}}{h} \right)^m  \right],
\label{eq: disjoin}
\end{align}
where two terms represent the short-range repulsive force and the long-range attractive force between liquid and solid, respectively. The parameter $A_H$ is the Hamaker constant, and $m,n,k$ are dimensionless constants with $n > m > 1$, representing the strength of the repulsive and attractive force, and surfactant-induced attraction strength.
A thin precursor of thickness $h_{\infty}$ is assumed.

Let $\bT$ be the total stress tensor in the fluid
\begin{equation} \label{eq: totaltensor}
\bT = -\left[ p \right] \bI + \mu\left( \nabla \bm{u} + \nabla \bm{u}^\intercal \right),
\end{equation}
where $\left[ ( \cdot ) \right]$ denotes the jump, ``outside-inside'', of $( \cdot )$ across the boundary. Let $\bm{n}=(-h_x,1)/(1+h_x^2)^{1/2}$ be the local outer normal vector and $\bm{t}=(1,h_x)/(1+h_x^2)^{1/2}$ be the tangential vector. The stress tensor satisfies the stress balance equation at the interface in normal and tangential directions,
\begin{equation}
\begin{aligned}
	 \bm{n} \cdot \bT \cdot \bm{n} &= \sigma \left( \nabla \cdot \bm{n} \right), \\
	 \bm{n} \cdot \bT \cdot \bm{t} &= \nabla \sigma \cdot \bm{t}, 
\end{aligned}
\label{eq: nbalance}
\end{equation}
where the normal stress is balanced by the local curvature pressure associated with the surface tension $\sigma$ and the tangential stress is balanced by the local surface tension gradient. 
We employ the Langmuir equation of state to relate the interfacial tension and surfactant concentration, which is
\begin{equation}
\sigma = \sigma_0 \left[ 1+\beta \ln \left(1-\frac{\Gamma}{\Gamma_{\infty}}\right)\right]\,,
\label{eq: logsigma}
\end{equation}
where $\sigma_0$ is the surface tension without surfactants, 
$\beta$ is the surface elasticity number and $\Gamma_\infty$ is the maximum surfactant packing \cite{ervik2018influence}.

Electro-dewetting is a dissipative process that occurs with an electric current and electrochemical reactions in the droplet. Since ionic surfactants are usually stable in electrochemical reactions \cite{vittal2006beneficial}, we assume surfactant ions do not participate in the electrochemical reactions. Hence, surfactant transport on the liquid-air interface is only driven by convection, diffusion, and electrophoresis \cite{stone1990simple,oron1997long}. 
The transport process on the interface is governed by the Nernst--Planck equation
\begin{align}\label{eq: gammatransport}
\frac{\partial \Gamma}{\partial t} + \nabla_{\!s}\cdot \left( \Gamma \bm{u}_{s}\right)  
&= D_{s}\nabla_{\!s}^2\Gamma - D_{e} \nabla_{\!s} \left( \Gamma \bE_s\right),
\end{align}
where $\nabla_{\!s}$ is the surface gradient operator, $\nabla_{\!s}=(\bI-\bm{nn})\cdot\nabla$, $\bm{u}_{s}$ is the velocity along the surface, $\bm{u}_{s}=(\bI-\bm{nn})\cdot\bm{u}$, $D_{s}$ is the diffusion coefficient on the air-liquid surface, $\bE_s$ is the tangential component of the electric field $\bE$, $\bE_s=(\bI-\bm{nn})\cdot\bE$, and $D_e=(zeD_s)/(k_B T)$.

Since an electric current flows to maintain the electric field in the electrically conductive droplet, electro-dewetting is accompanied by electrochemical reactions with hydroxide ions as the main charge carrier. In contrast, the long-chain surfactant ions used for electro-dewetting (\eg, DTA$+$) do not participate in the electrochemical reaction and, thus, do not create any current. Indeed, this type of ionic surfactant is widely used in electrocatalysis for electrode surface modification, which infers its stability in most electrochemical reactions \cite{vittal2006beneficial}. We further assume a uniform concentration of charge carriers (\ie, hydroxide ions) and electroneutrality within the droplet \cite{newman2021electrochemical}, and that free charge from surfactants does not affect the electric field distribution.
The constant voltage applied to the droplet leads to a steady current, in which the electric current density $J$ is described by 
\begin{equation}\label{eq: j1}
    \nabla \cdot J = 0.
\end{equation}
In addition, the electric current density $J$ is related to the electric field $\bE$ by
\begin{equation}\label{eq: j2}
    J = \kappa \bE,
\end{equation}
where $\kappa$ is the electric conductivity of the electrolyte. Since the steady current does not induce a time-varying magnetic field, one can define a potential function $\psi$ such that $\bE=-\nabla \psi$. From Equations \eqref{eq: j1} and \eqref{eq: j2}, we know that $\psi$ satisfies the Laplace equation 
\begin{equation}
\begin{aligned}
\nabla^{2}\psi = 0.
\end{aligned}
\label{eq: efield}
\end{equation}

At the free surface, the tangential electric field is continuous while the normal electric field has a jump due to the displacement current \cite{lopez2011charge}, such that
\begin{equation}
\left[ \varepsilon_{\sigma}\bE\right] \cdot \bm{n} = \frac{\partial q}{\partial t}, \quad \left[ \bE \right] \cdot \bm{t}=0,
\label{eq: balanceE}
\end{equation}
where $\varepsilon_{\sigma}$ is the electric conductivity.
The electric field jump in the normal direction is proportional to the charge relaxation time and is negligible $({\partial q}/{\partial t}\approx 0)$ for smaller drops \cite{nganguia2019effects}.

\subsection{Non-dimensionalization and model simplification}\label{sec: ndbd}
Let $H$ and $L$ be the characteristic thickness and length of a droplet resting on a substrate without an external field. The equations presented above are rendered dimensionless by adopting the following scalings,
\begin{equation}
\begin{array}{lll}
x = L\hx,         & \quad (y,h) = H(\hy,\hh),                         & \quad (u,v)=U(\hu, \delta\hv), \\
t=(L/U)\hat{t},   & \quad (p,\Pi) =(\mu UL/H^2)(\hat{p},\hat{\Pi}),   & \quad 
(\psi,\phi)=\Psi_0(\hat{\psi}, \hat{\phi}),\\
 (\Gamma,\Gamma_\infty) = \Gamma_0(\hat{\Gamma},\hat{\Gamma}_\infty), & \quad 
(\sigma_0,\sigma) = \left(\mu UL/H\right)(\hat{\sigma}_0,\hat{\sigma}), & \\
\end{array}
\end{equation}
where $\delta=H/L \ll 1$ is the lubrication parameter, $\Psi_0$ is the maximum external potential, $\Gamma_0$ is the characteristic surfactant concentration on the interface.
In the following, let us drop the $\hat{\cdot}$ notation for dimensionless equations.
We also use the subscript notation for $x$, $y$, and $t$ to indicate partial derivations.

After combining equations~\eqref{eq: ns}\eqref{eq: totaltensor}\eqref{eq: nbalance} and \eqref{eq: logsigma} and dropping lower order terms in the low Reynolds number limit with $\delta\to 0$, one obtains the following equations for the liquid domain $\Omega_L$,
\begin{equation} \label{eq: uyy}
    \begin{aligned}
    u_{yy} &=\left( p-\Pi \right)_{x},\qquad
u_{y}  &= -\sigma_0\beta\frac{\Gamma_x}{\Gamma_\infty-\Gamma}, 
\end{aligned}
\end{equation}
and the pressure $p$ at air-liquid interface $y=h(x,t)$ satisfies
\begin{equation}\label{eq: pressure}
    \begin{aligned}
p &=p_{G}-h_{xx}\sigma_0\left[1+\beta\ln\left(1-\frac{\Gamma}{\Gamma_\infty}\right)\right],
\end{aligned}
\end{equation}
where $p_{G}$ is the atmospheric pressure and $\Pi$ is the disjoining pressure. By integrating equation \eqref{eq: uyy} with the no-slip boundary condition that $u(x,0)=0$, we obtain
\begin{align}
u(x,y) = \left(p-\Pi\right)_x \left(\frac12y^{2}-hy\right) - \left( \frac{\sigma_0\beta \Gamma_x}{\Gamma_\infty-\Gamma}\right)y.
\end{align}
Further with the incompressible condition \eqref{eq: ns} and the kinematic free surface condition \eqref{eq: freebd}, we have the governing equation for the film thickness,
\begin{align}
h_t - \left[ \frac{h^3}{3}\left(p-\Pi\right)_x + \frac{h^2}{2}\left(\frac{\sigma_0\beta \Gamma_x}{\Gamma_\infty-\Gamma}\right)\right]_x &= 0,
\label{eq: thickness}
\end{align}
where the pressure $p$ is given in equation \eqref{eq: pressure} and $\Pi$ is the dimensionless disjoining pressure computed from equation \eqref{eq: disjoin},
\begin{equation} \label{dimless_pi}
    \Pi(h,\Gamma) = \frac{B}{h_{\infty}^n} \left[ \left( \frac{h_{\infty}}{h} \right)^n - \left(1+\frac{\Gamma}{\Gamma_\infty}\right)^k\left( \frac{h_{\infty}}{h} \right)^m  \right],
\end{equation}
with $B=A_H/6\pi\mu ULH^{n-2}$.
Similarly, we apply the lubrication analysis and rescaling to \eqref{eq: gammatransport} and obtain a reduced transport equation for the surfactant concentration,
\begin{align} \label{eq: gammaeq}
    \Gamma_t + (\Gamma u)_x  = \frac{1}{\text{Pe}}\Gamma_{xx} - D  \left[ \Gamma \left( \psi_x + h_x\psi_y \right)\right]_x
    \,,
\end{align}
where $\text{Pe}=L^2/D_sT$ is a Peclet number, representing the interaction between transport due to diffusion in the gas and the advective transport in the fluid layer, and $D=\Psi_0/\text{Pe}$ is the electric field induced diffusion coefficient.
With the thin film assumption $\delta=H/L \ll 1$, one can drop lower order terms in equation \eqref{eq: efield} and obtain
\begin{equation}
\psi(x,y,t) = \Psi_1(x,t)y + \Psi_0(x,t), \quad (x,y)\in\Omega_L,
\label{eq: ppsi}
\end{equation}
where $\Psi_0$ are $\Psi_1$ can be determined by the external potential imposed on the boundary. We will discuss the selection of their forms in the following subsection for two different electrode configurations, which correspond to droplet deformation and directional movement.

With the electric fields in place, we come to a closed form of equations for the film thickness $h(x,t)$ and surfactant concentration $\Gamma(x,t)$ to describe the droplet actuation process under versatile electric manipulation,
\begin{subequations}
\begin{equation}
h_t - \left[\frac{h^3}{3}\tilde{p}_x + \frac{h^2}{2}\left(\frac{\sigma_0\beta \Gamma_x}{\Gamma_\infty-\Gamma}\right)\right]_x = 0,
\label{eq: h_eqn}
\end{equation}
\begin{equation}
\Gamma_t - \left[ \frac{h^2\Gamma}{2}\tilde{p}_x + {h\Gamma}\left(\frac{\sigma_0\beta \Gamma_x}{\Gamma_\infty-\Gamma}\right)\right]_x = {\textstyle\frac{1}{\text{Pe}}}\Gamma_{xx} - D\left[ \Gamma \mathcal{E}(x,h,t) \right]_x,
\label{eq: gamma_eqn}
\end{equation}
where the pressure $\tilde{p}$ incorporates both the intermolecular force by $\Pi(h,\Gamma)$ and the surface tension,
\begin{equation}
    \tilde{p} = -\Pi(h,\Gamma)-\sigma_0h_{xx}\left[1+\beta\ln\left(1-\frac{\Gamma}{\Gamma_\infty}\right)\right].
\end{equation}
\label{eq:main}
\end{subequations}
The last term in equation \eqref{eq: gamma_eqn} characterizes the electric effects on the surfactant dynamics, where $\mathcal{E}(x,h,t) = \frac{\partial}{\partial x}\psi(x,h(x,t),t)$ represents the $x$ component of the electric field at the free surface $h(x,t)$. 
The approximation of the electric potential at $h(x,t)$, $\psi(x,h,t)\approx \Psi_1h+\Psi_0$,  will be discussed in Section \ref{sec:electric_field}.

\subsection{Electric field model} \label{sec:electric_field}
We obtain the formula of $\Psi_0(x,t)$ directly from the electric potential imposed on the substrate, leaving $\Psi_1(x,t)$ to be determined. 
Instead of solving Equation \eqref{eq: efield} with time-dependent boundary conditions \eqref{eq: balanceE} of the entire domain, we propose a formula for $\Psi_1(x,t)$ such that the approximated electric field is close to the true electric field at the air-liquid interface, where the thin film equation is defined. In particular, we have
\begin{equation}
    \psi(x,h(x,t),t) \approx \Psi_1(x,t)h(x,t)+\Psi_0(x,t).
\label{eq:psi_approx}
\end{equation} 
We provide the formulas of $\Psi_0$ and $\Psi_1$ in terms of two actuation configurations that are applied in electro-dewetting experiments.

\subsubsection{Deformation of a droplet}
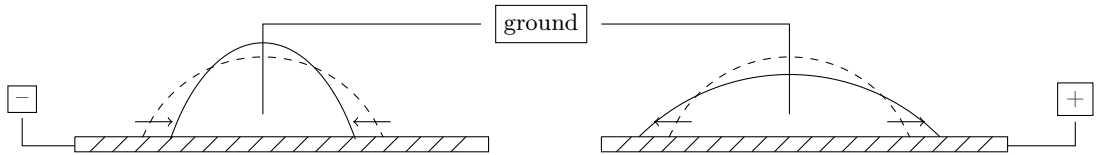
\begin{figure}[htpb] 
\begin{center}
\begin{tikzpicture}[scale=1]
[roundnode/.style={circle, draw=black!60, fill=black!5, very thick, minimum size=7mm}]
\draw[-] (-2.5,1) -- (3,1) -- (3,0.8) -- (-2.5,0.8) -- (-2.5,1); 
\foreach \x in {-2.1, -1.8, ..., 2.9}
  	\draw[-] (\x,1) -- (\x-0.2,0.8);
\draw[-] (4.5,1) -- (9.9,1) -- (9.9,0.8) -- (4.5,0.8) -- (4.5,1); 
\foreach \x in {4.9, 5.2, ..., 9.9}
  	\draw[-] (\x,1) -- (\x-0.2,0.8);
\draw[-] (-2.5,0.88) -- (-3.2,0.88) -- (-3.2,1.25); 
\draw[-] (9.9,0.88) -- (10.8,0.88) -- (10.8,1.2); 

\draw[-] (0,1.3) -- (0,2.5) -- (2.9,2.5); 
\draw[-] (7,1.3) -- (7,2.5) -- (4.5,2.5); 

\node[draw] at (-3.2, 1.5)  {--}; 
\node[draw] at (10.8, 1.5)  {+}; 
\node[draw] at (3.7, 2.5)  {ground}; 

\draw[scale=1,dashed,domain=-1.6:1.6,smooth,variable=\x] plot (\x,{sqrt(3-\x*\x)+0.33}); 
\draw[scale=1,domain=-1.23:1.23,smooth,variable=\x] plot (\x,{sqrt(9-4*\x*\x)-0.75});

\draw [->] (-1.7,1.2) -- (-1.2,1.2);
\draw [<-] (1.2,1.2) -- (1.7,1.2);

\draw[scale=1,dashed,domain=-1.6:1.6,smooth,variable=\x] plot (\x+7,{sqrt(3-\x*\x)+0.33}); 
\draw[scale=1,domain=-2:2,smooth,variable=\x] plot (\x+7,{sqrt(8-\x*\x)-1});

\draw [->] (5.7,1.2) -- (5.2,1.2);
\draw [->] (8.3,1.2) -- (8.8,1.2);
\end{tikzpicture}

\end{center}
\caption{Schematic diagrams of the electro-dewetting and electro-rewetting experiments. At steady states,  a droplet containing cationic (\ie, positively charged) surfactants sits on a bare silicon (indicated by the dashed line). A pin electrode is inserted into the droplet from the top with a ground (zero) potential. 
When a negative potential is applied to the substrate electrode (left figure), cationic surfactants are attracted and absorbed on the substrate, and the droplet beads up (\ie, dewets).
When a positive potential is applied to the substrate (right figure), cationic surfactants are attracted away and desorbed from the substrate, and the droplet spreads again (\ie, rewets).}
\label{fig: case1}
\end{figure}

We first consider a pin configuration where a sessile droplet deforms according to the external potential, applied by a pin electrode and a substrate electrode. Fig.~\ref{fig: case1} depicts the electro-dewetting and electro-rewetting experiment of a sessile droplet.
We assume that the pin electrode always imposes a zero potential at the center top of the droplet as the droplet spreads and beads up. The substrate electrode has a homogeneous (uniform) potential. 
To model the electric potential in the sessile droplet, we assume the droplet center is at $x=0$, which yields $\Psi_0(x,t)=\Psi(t)$ and $\Psi_1(0,t)h(0,t)=0$, where $\Psi(t)$ is the homogeneous substrate potential. As we increase the potential $\Psi(t)$, the potential $\psi(x, h(x,t),t)$ should increase proportionally as well. We propose an approximation of $\Psi_1$ as 
\begin{equation} \label{eq: pin_electrode}
    \Psi_1(x,t) = A(x)\Psi(t), \quad A(x) = -\exp(-cx^2)/h(0,t).
\end{equation}
This approximation is motivated by numerical observations that the electric potential decays exponentially on the interface, and $c$ is a dimensionless rescaling parameter related to the droplet width and can be determined by interpolation.

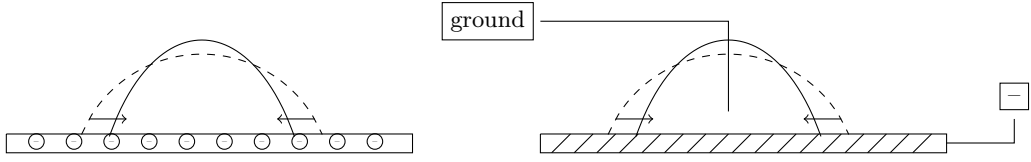
\begin{figure}[htpb]
\centering
\begin{tikzpicture}[scale=1]
[roundnode/.style={circle, draw=black!60, fill=black!5, very thick, minimum size=7mm}]
\draw[-] (-2.6,1) -- (2.8,1) -- (2.8,0.75) -- (-2.6,0.75) -- (-2.6,1); 
\foreach \x in {-2.2, -1.7, ..., 2.4}
  	\node[draw,circle, scale=0.4] at (\x,0.9) {--};

\draw[-] (4.5,1) -- (9.9,1) -- (9.9,0.75) -- (4.5,0.75) -- (4.5,1); 
\foreach \x in {4.9, 5.2, ..., 9.9}
  	\draw[-] (\x,1) -- (\x-0.25,0.75);
\draw[-] (9.9,0.88) -- (10.8,0.88) -- (10.8,1.2); 

\draw[-] (7,1.3) -- (7,2.5) -- (4.5,2.5);

\node[draw] at (10.8, 1.5)  {--}; 
\node[draw] at (3.8, 2.5)  {ground}; 

\draw[scale=1,dashed,domain=-1.6:1.6,smooth,variable=\x] plot (\x,{sqrt(3-\x*\x)+0.33}); 
\draw[scale=1,domain=-1.23:1.23,smooth,variable=\x] plot (\x,{sqrt(9-4*\x*\x)-0.75});

\draw [<-] (-1,1.2) -- (-1.5,1.2);
\draw [<-] (1,1.2) -- (1.5,1.2);

\draw[scale=1,dashed,domain=-1.6:1.6,smooth,variable=\x] plot (\x+7,{sqrt(3-\x*\x)+0.33}); 
\draw[scale=1,domain=-1.23:1.23,smooth,variable=\x] plot (\x+7,{sqrt(9-4*\x*\x)-0.75});

\draw [<-] (6,1.2) -- (5.5,1.2);
\draw [<-] (8,1.2) -- (8.5,1.2);
\end{tikzpicture}
\caption{A schematic diagram of real autophobing mechanism and the proposed alternative. Left: droplet containing cationic surfactants automatically beads up after being placed on a substrate with negative surface charge, as indicated by the arrow directions. Right: a virtual potential is used to mimic the effect of surface charge for autophobing.}
\label{fig: case3}
\end{figure}

This electric field approximation \eqref{eq: pin_electrode} also applies to the case of \emph{autophobing}. Autophobing is a phenomenon that a droplet containing ionic surfactants automatically beads up (\ie, dewets) after spreading on a surface \cite{hare1955autophobic}. Craster and Matar \cite{craster2007autophobing} developed a model to explain autophobing effects for liquid ladened with soluble surfactants and considered the case when the bulk concentration is above the critical micelle concentration (cmc).
Instead of considering the dynamics of bulk surfactants, we assume the surfactants are insoluble.
When an aqueous droplet containing cationic surfactants rests on a glass substrate, the surface is charged negatively in contact with an aqueous liquid, which shares a similar underlying mechanism as electro-dewetting. 
Fig.~\ref{fig: case3} (left) schematically illustrates the autophobing effect of a droplet.
For a water droplet of neutral pH values, the negative surface charges attract cationic surfactants to the substrate near triple points, and the droplet beads up (\ie, dewets). 
We introduce a virtual electric potential formed by a pin electrode and a substrate electrode as shown in Fig.~\ref{fig: case3} (right), where the electric potential also takes the form in equation~\eqref{eq: pin_electrode}. Despite the departure from the actual physics, which involves the electric field within the electric double layer on the substrate surface, this virtual bias applied across the droplet provides a convenient method to mimic the autophobing.

\begin{figure}[ht]
    \includegraphics[width=0.42\textwidth]{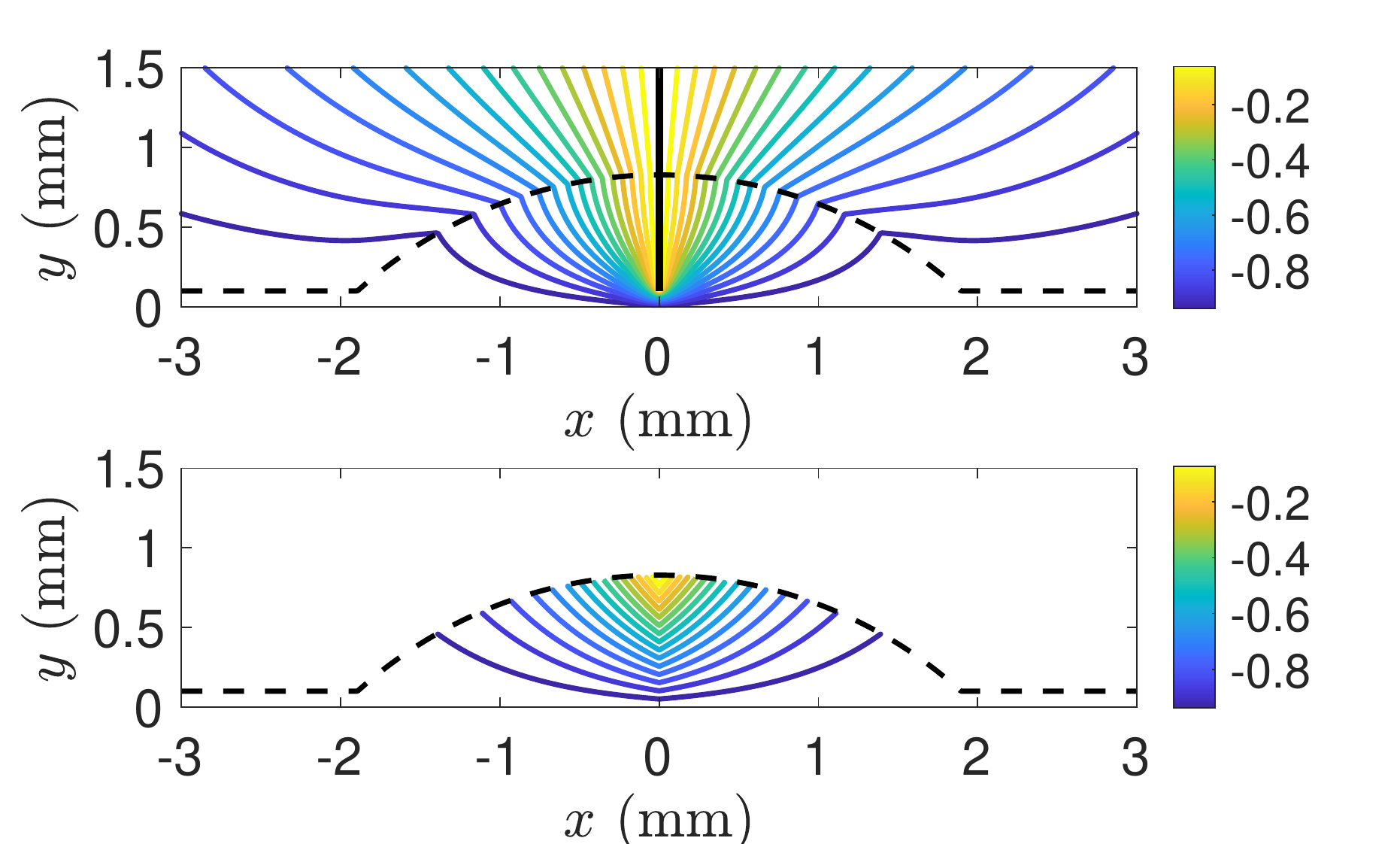}
    \includegraphics[width=0.42\textwidth]{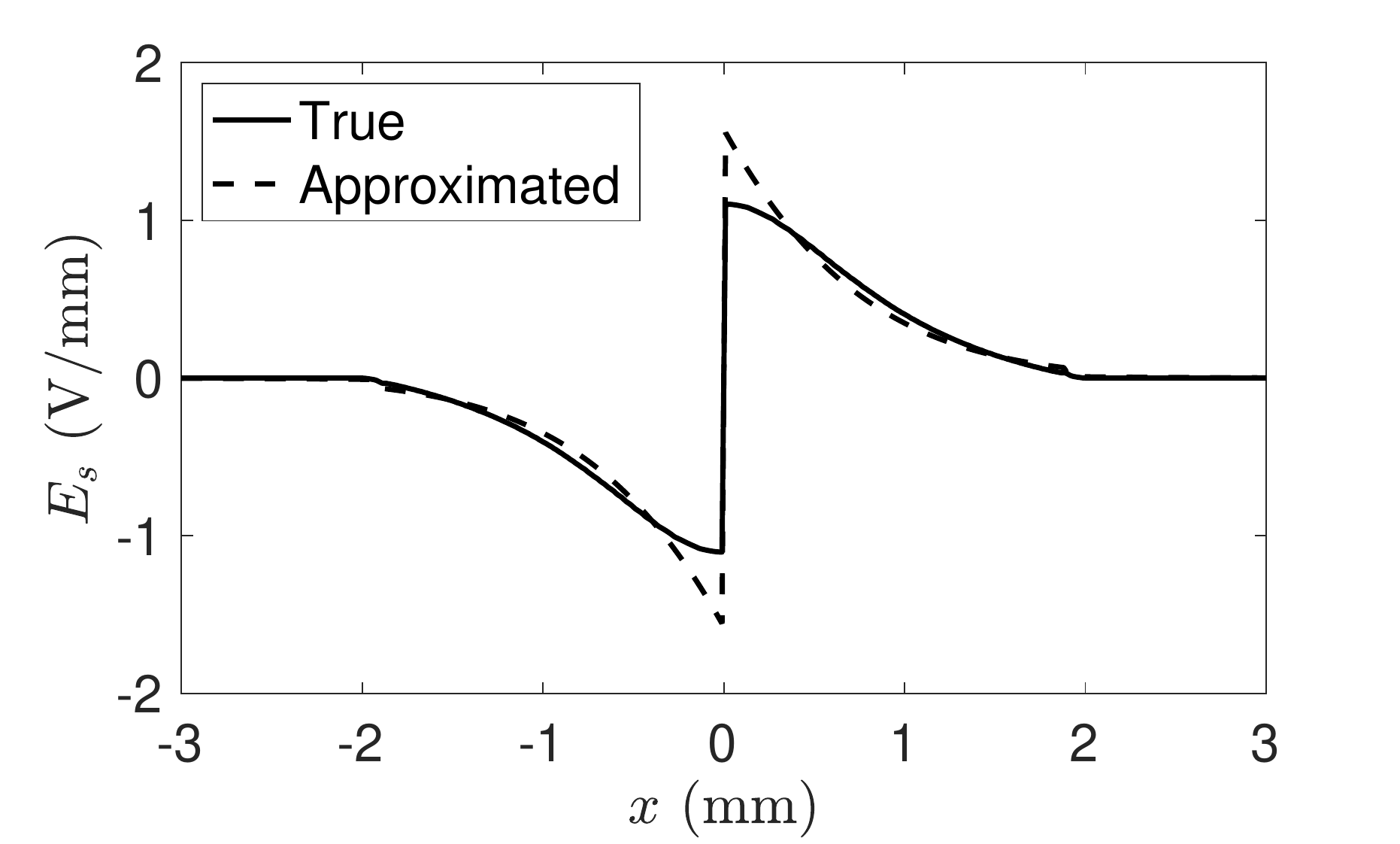}
    \caption{\label{fig: electricfield}(Color online) The left two figures show the electric equi-potential lines for true and approximated electric potential values. The droplet shape is indicated by dashed lines. Left top: electric equi-potential lines with pin potential $\psi_\text{pin}=0$ at $x=0, y\ge0.1$ (marked in black thick solid line) and a homogeneous potential equal to $-1$ at $y=0$. Left bottom: equi-potential lines computed by the proposed formula in equations~\eqref{eq: pin_electrode}. The parameter $c=1.5767$ is obtained from a two-point interpolation of electric potential at $\psi(x_1,y_1)=0$ and $\psi(x_2,y_2)=-0.999$, where $x_1=0, y_1=h(x_1)$ corresponds to the zero potential from the pin electrode and $x_2=1.8947, y_2=h(x_2)$ corresponds to the triple point on the right. Right: a comparison of the tangential electric field strength between true and approximated values. The electric field is not continuous due to the singularity caused by the pin. The difference near the pin electrode is inconsequential to our interest, which is around the triple points.}
\end{figure}

To the best of our knowledge, this is the first use of lubrication theory to approximate the electric potential in a pin-electrode open configuration. To validate the model in \eqref{eq: pin_electrode}, we compare the real electric field to the proposed ones for the two cases corresponding to our experiments.
The real electric field is obtained by solving equation~\eqref{eq: efield} in a 2-dimensional domain with the five-point stencil finite difference method. We consider a large enough computational domain $[-10,10]\times[0,10]$ and consider the electric distribution near the droplet region, where the droplet profile is approximated by
\begin{equation}
    h(x)=\max\{\sqrt{\max(8-x^2, 0)}-2, 0.1\}.
\end{equation}
We apply a Dirichlet boundary condition on the domain boundary and the pin electrode position. We assume that the electric field only affects the dynamics of surfactants, which exist only on the water-air interface. We further compute the electric field strength $E_s$ on the water-air interface, which is
\begin{equation}
E_s=\psi_x(x,y)+h_x(x)\psi_y(x,y)|_{y=h(x)}.\end{equation}
In Fig.~\ref{fig: electricfield}, we compare the approximated results using the formula in equation~\eqref{eq: pin_electrode} with the true solutions from equation~\eqref{eq: efield} for the pin electrode configuration. 

\subsubsection{Shifting of a droplet}
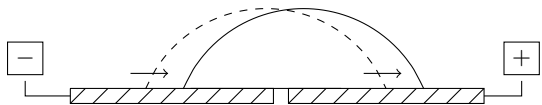
\begin{figure}[htpb] 
\begin{center}
\begin{tikzpicture}
\draw[-] (-2.6,1) -- (0.1,1) -- (0.1,0.8) -- (-2.6,0.8) -- (-2.6,1); 
\draw[-] (0.3,1) -- (2.9,1) -- (2.9,0.8) -- (0.3,0.8) -- (0.3,1); 
\foreach \x in {-2.4, -2.1, ..., 0, 0.5, 0.8, ..., 3}
  	\draw[-] (\x,1) -- (\x-0.2,0.8);
\node[draw] at (-3.2, 1.4) (a) {$-$};
\node[draw] at (3.4, 1.4) (a) {$+$};
\draw[-] (-1,1) -- (1,1); 
  
\draw [-] (-2.6,0.9) -- (-3.2, 0.9) -- (-3.2, 1.1);
\draw [-] (2.9,0.9) -- (3.4,0.9) -- (3.4,1.1);

\draw[scale=1,dashed,domain=-1.6:1.6,smooth,variable=\x] plot (\x,{sqrt(3-\x*\x)+0.33}); 
\draw[scale=1,domain=-1.1:2.1,smooth,variable=\x] plot (\x,{sqrt(3-(\x-0.5)*(\x-0.5))+0.33}); 

\draw [->] (-1.8,1.2) -- (-1.3,1.2);
\draw [->] (1.3,1.2) -- (1.8,1.2);

\end{tikzpicture}
\end{center}
\caption{A schematic diagram of droplet shifting by applying an electric potential between the two planar electrodes on the substrate. A droplet rests on its steady-state configuration (dashed line) initially. After imposing an external electric field, the droplet shifts towards the electrode with positive potential (as indicated by the arrows).
}
\label{fig: case2}
\end{figure}

\begin{figure}[htpb]
\centering
    \includegraphics[width=0.44\textwidth]{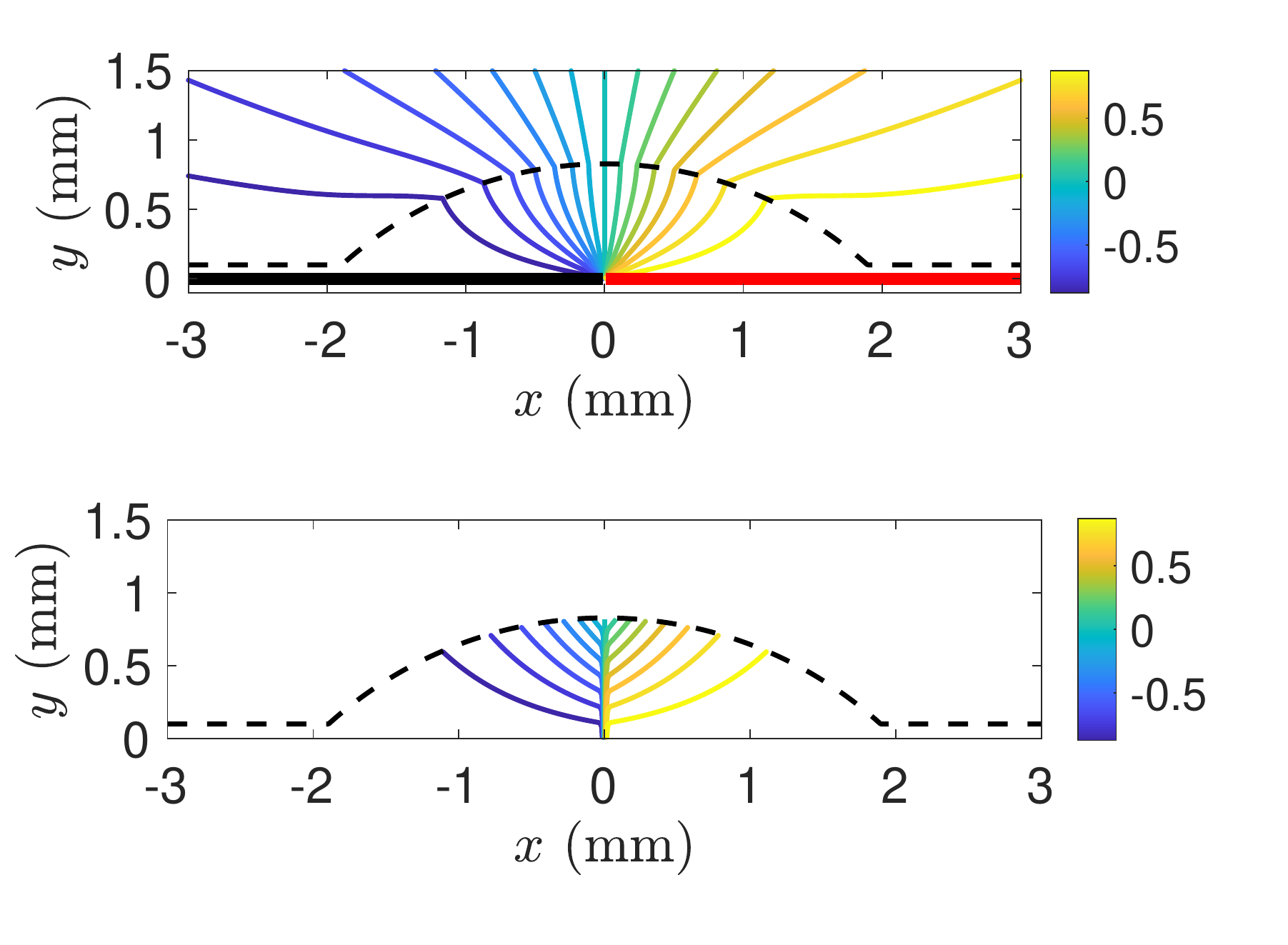}
    \includegraphics[width=0.44\textwidth]{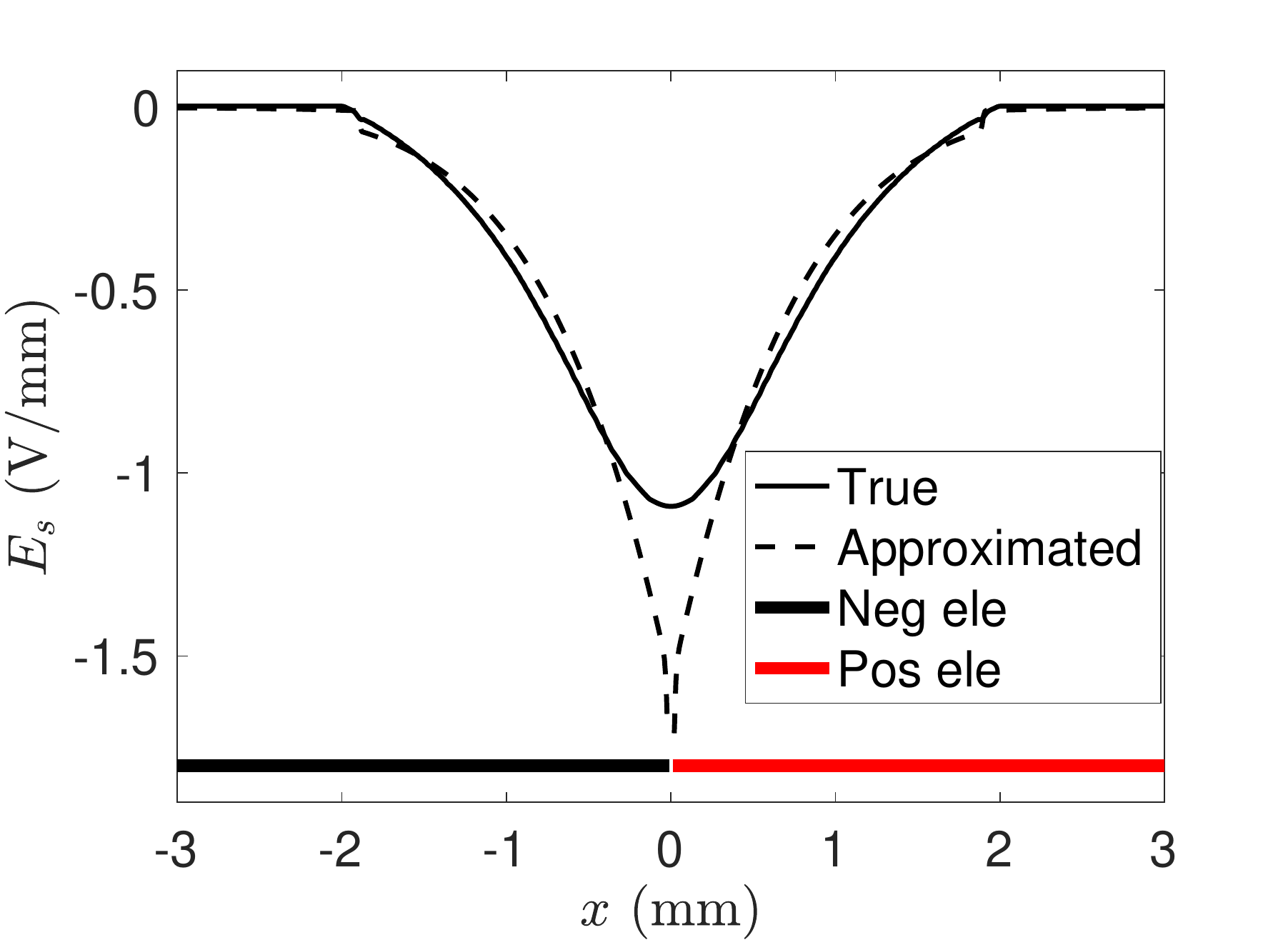}
    \caption{(Color online) The left two figures show the electric equi-potential lines for true and approximated potential. The droplet shape is indicated by dashed lines. Left top: electric equi-potential lines with substrate potential $\psi=-1$ for $x<0, y=0$ (marked in black thick solid line) and $\psi=1$ for $x>0, y=0$ (marked in red thick solid line). Left bottom: equi-potential lines computed by the proposed formula in equations~\eqref{eq: case2}. The parameter $c=1.5767$ is obtained from a two-point interpolation of electric potential at $\psi(x_1,y_1)=0$ and $\psi(x_2,y_2)=0.999$, where $x_1=0, y_1=h(x_1)$ corresponds to the jump and $x_2=1.8947, y_2=h(x_2)$ corresponds to the triple point on the right. Right: a comparison of the tangential electric field strength between true and approximated values, with electrode positions marked in thick lines. The true tangential electric field is continuous and reaches its maximum near the discontinuity jump location ($x=0$).}
    \label{fig: nonhomogeneousfiled}
\end{figure}

Droplet transport is one of the basic droplet operations for DMF and is experimentally demonstrated by using electro-dewetting \cite{li2019ionic}. We consider an open configuration where two planar electrodes are attached to the substrate and impose nonhomogeneous potential to induce directional transport as illustrated in Fig.~\ref{fig: case2}.
To simulate the electric potential in this configuration, we consider a simple case when the potential takes the form of a shifted sign function centered at $a(t)$, \ie, 
\begin{equation} \label{eq: case3_psi0}
    \Psi_0(x,t) = \Psi(t)\text{sgn}(x-a(t)) = \left\{
    \begin{array}{cl}
       -\Psi(t),  & x<a(t) \\
        0,  & x = a(t) \\
        \Psi(t),  & x > a(t)
    \end{array}
    \right.\,.
\end{equation}
Here, $a(t)$ is the location between two planar electrodes and may change over time.
In practice, we approximate $\Psi_0$ with a hyperbolic tangent function to avoid singularity at $x=a(t)$. Similarly, we propose an approximated form for $\Psi_1(x,t)$ as
\begin{equation} \label{eq: case2}
    \Psi_1(x,t) = \Psi_0(x,t)A(x,t) = -\Psi_0(x,t)\exp(-c(x-a(t))^2)/h(a(t),t)\,.
\end{equation}
Similarly, we report the electric potential and field when the substrate consists of two planar electrodes with the proposed approximation in equation~\eqref{eq: case2} and compare them with the true values in Fig.~\ref{fig: nonhomogeneousfiled}. The substrate potential has a jump at $x=0$ and the potential takes the form of $\Psi_0(x) = \tanh({100x})$.

\section{Experimental measures} \label{sec: experiments}
\begin{figure}[htp]
    \includegraphics[width=0.5\textwidth]{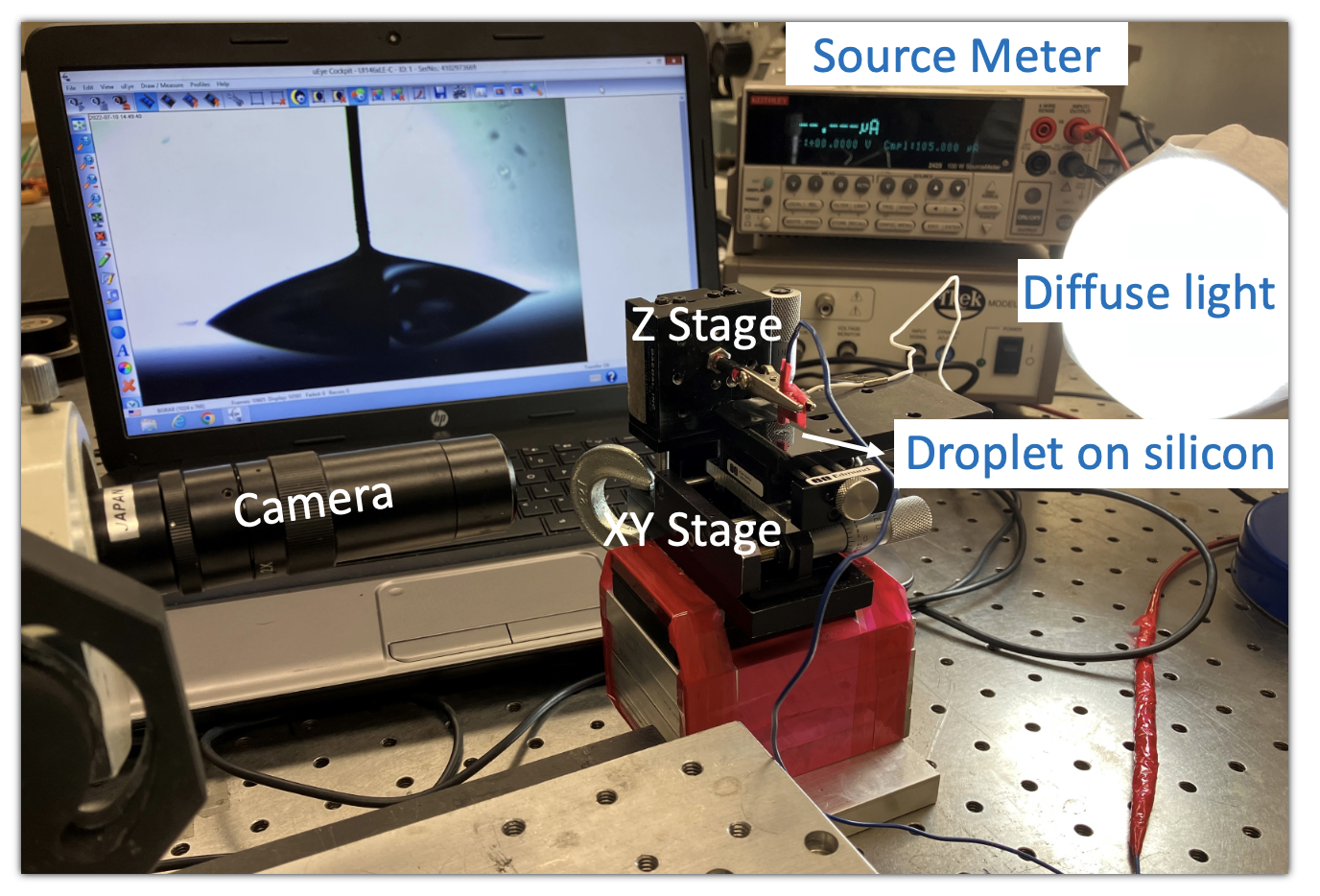}
    \caption{A picture of the experimental setup to conduct electro-dewetting experiment and measure contact angle of a sessile droplet.}
    \label{fig: experiment}
\end{figure}

Following Li et al. \cite{li2019ionic}, electro-dewetting experiments are performed for this study to provide the contact angle data as a comparison for numerical results in Section \ref{sec: numerics}, aiming to help assess the proposed model. As shown in the test setup of Fig.~\ref{fig: experiment}, a droplet rests on a piece of highly doped, thus conductive, silicon wafer, which functions as a substrate electrode. 
A platinum wire (100-$\mu$m diameter), which functions as a pin electrode, is vertically inserted into the droplet from the top. The silicon piece is placed on an XY stage, and the wire is mounted on an independent Z stage. Both the substrate and pin electrodes are connected to a source meter (Keithley 2425 Source Meter). A camera (Edmund Optics IDS Imaging U3-3040SE) with a long-range lens (OPTEM 125) is mounted on an XYZ stage to record the side view of the droplet under electric actuation, from which the contact angle data is obtained using an image analysis software (ImageJ with the DropSnake plugin). The droplets used for the experiments are fresh $0.2$ mM DTAB solution, which is prepared by dissolving DTAB surfactant powder (Sigma Aldrich) into deionized water. The pH of the solution is measured to be $6.5$ by an electronic pH meter (Apera PH60). 

Before the test, the silicon piece is thoroughly rinsed by deionized water and blow-dried by nitrogen gas to obtain a clean surface. After pipetting a droplet (about $3~\mu$L) of surfactant solution onto the silicon piece, we adjust the XY stage to center the droplet beneath the wire and then lower the Z stage to insert the wire into the droplet. The tip of the wire maintains $85~\mu$m above the silicon surface throughout the testing. The real-time camera view is used to facilitate accurate positioning of the droplet and the wire. 
The source meter provides $-3~\mu$A or $+3~\mu$A current to the substrate electrode for the electro-dewetting or electro-rewetting actuation, respectively, which results in an external voltage of around $-3$ V or $+4$ V for the current experimental setup. Since the solution is electrically conductive, an electric current flows through the droplet to maintain an electric field that induces electro-dewetting or electro-rewetting \cite{li2019ionic}. Although one can use any electrical power source in principle, a current source is more convenient to apply a desired voltage across the thin and conductive droplet (\ie, a small resistor). The current is electrically connected in series to other components (\eg, probe, silicon pieces, cables) in an experimental setup.

During the experiment, we observe that, after dispensed on the silicon surface, the droplet first spreads until its contact angle decreases to 18.5$^{\circ}$ and then retracts by autophobing until its contact angle reaches 25.7$^{\circ}$. 
Note that the contact angles of a droplet reported here are the mean values of the left and right contact angles. After the droplet settles to a steady state, an electric potential of $-3$ V is applied to the substrate electrode to induce electro-dewetting, which increases the contact angle to $59.2^{\circ}$. Then, a potential of $+4$ V is applied to induce electro-rewetting, which decreases the contact angle to $14.5^{\circ}$. The pin electrode remains grounded for the electro-dewetting and electro-rewetting actuations in this study.

\section{Numerical Studies}\label{sec: numerics}
To numerically solve the governing model, we use the Keller box method \cite{keller1971new} to decompose the coupled fourth-order PDE system \eqref{eq:main} into a system of first-order differential equations,
\begin{equation}
\begin{split}
    & k = h_x,\quad p =-\Pi(h,\Gamma)-\sigma_0k_x\left[1+\beta\ln(1-\Gamma/\Gamma_{\infty})\right],\quad  q = h^2p_x, \quad w = \Gamma_x,\\
    & 
    h_t = \left(\frac{hq}{3}+ \frac{h^2w}{2}\frac{\sigma_0\beta }{\Gamma_{\infty}-\Gamma}\right)_x,\quad  \Gamma_t = \left(\frac{\Gamma q}{2}+\frac{h\Gamma w\sigma_0\beta}{\Gamma_{\infty}-\Gamma}\right)_x +{\textstyle\frac{1}{\text{Pe}}}w_x - D(\Gamma\mathcal{E})_x,
  \end{split}
\label{eqn:numeric_main}
\end{equation}
where the disjoining pressure $\Pi(h,\Gamma)$ takes the form in equation~\eqref{dimless_pi}. The last term in equation~\eqref{eqn:numeric_main} originates from the $x$ component of the electric field $\mathcal{E} = \partial_x\psi(x,h(x,t),t)$ at the interface $y = h(x,t)$, where the electric potential $\psi(x,h,t)$ is approximated by equation~\eqref{eq:psi_approx}. 
This system is then solved by fully implicit second-order centered finite differences over the domain $-L \le x \le L$. 
We impose no-flux boundary conditions on both the film thickness $h$ and the surfactant concentration $\Gamma$, $h_x = h_{xxx} = \Gamma_x = 0$, at $x =\pm L$. This set of boundary conditions guarantee the mass conservation of the fluid in time. We also impose $\psi_x = 0$ at the boundary $x=\pm L$ in order to guarantee a conserved quantity of surfactants over the domain. 
Table \ref{tab: nomenclature} in the Appendix shows the relevant nomenclature and corresponding value ranges for the experiments.
For the rest of the paper, we specify the dimensionless system parameters as
\begin{equation*}
\begin{aligned}
    & h_{\infty}=0.01, \quad  \Gamma_{\infty} = 10, \quad D = 0.005, \quad \text{Pe}=5000, \quad \sigma_0 = 0.0064, \\
    & \delta=0.2, \quad \beta = 0.023, \quad n=4, \quad m=3, \quad k=3,\quad B=3\times 10^{-8}.
\end{aligned}
\end{equation*}


\subsection{Relaxation to the unactuated state} \label{sec: relaxation}
\begin{figure}[htpb]
    \includegraphics[width=0.42\textwidth]{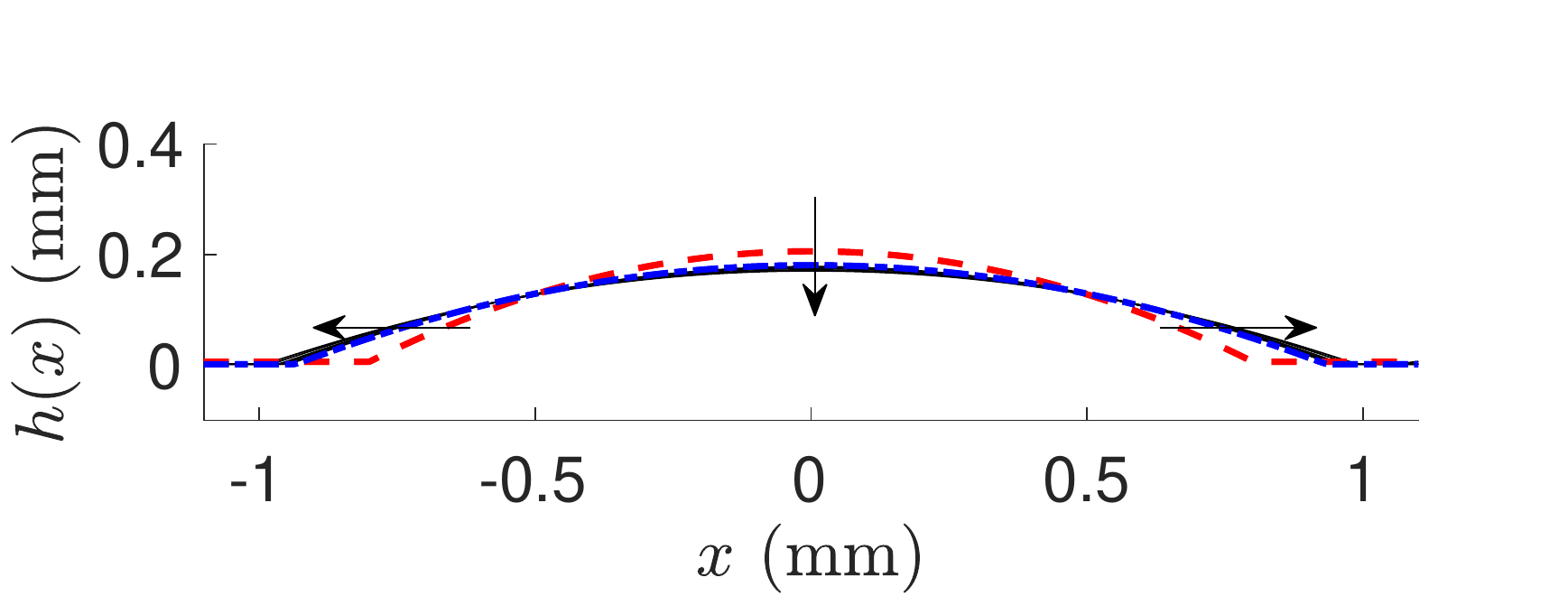} 
    \includegraphics[width=0.5\textwidth]{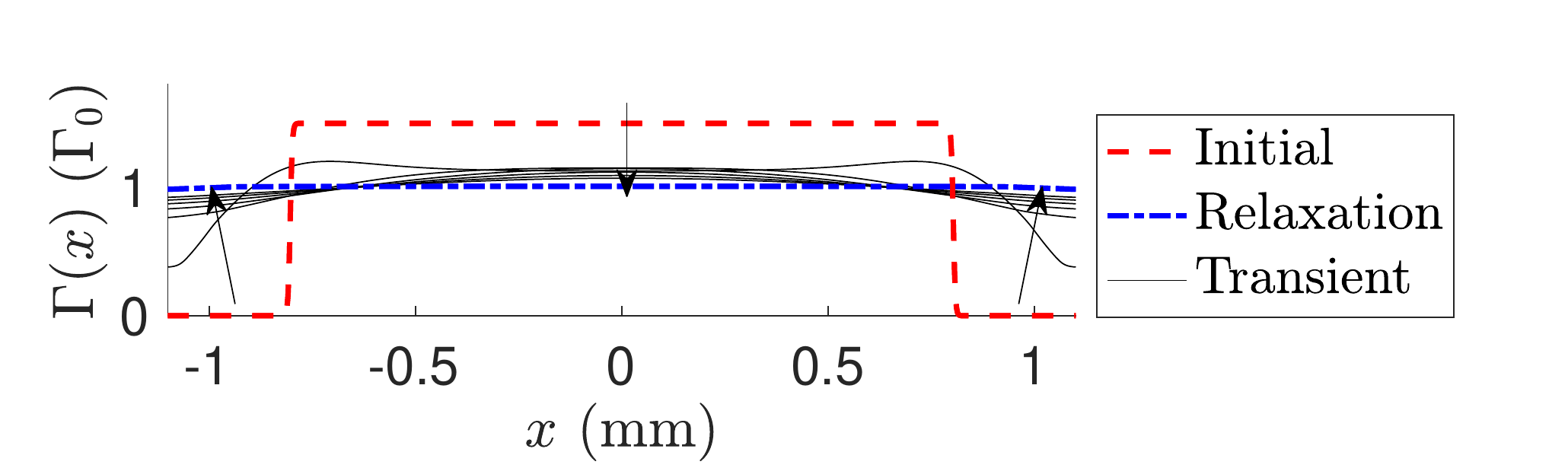} 
    \caption{The dynamics of a droplet arriving from an \emph{arbitrarily} chosen initial shape to its steady state without external electric field. The left picture shows the transient droplet profiles; while the right picture shows the dynamics of surfactant concentration ($\Gamma_0=\text{4.56e-2}~\text{mmol}/\text{m}^2$), which is estimated from the the soluble solution concentration (0.015cmc) times the characteristic length in the y-direction ($H$). The arrows show the trend of the evolution as time increases. The red dashed curves show the initial profiles of the film thickness $h$ and surfactant concentration $\Gamma$ at $t=0$s, and the blue dashed-dotted curves show the steady-state profiles of $h$ and $\Gamma$ at time $t=5$s.
    }
    \label{fig: relaxation}
\end{figure}

To rule out the influence of artificial initial conditions before the drop starts being strongly influenced by the electric field, we solve the model without the presence of external potential (\ie, $\Psi_0=\Psi_1=0$). We present the dynamics of film thickness $h$ and surfactant concentration $\Gamma$ in Fig.~\ref{fig: relaxation}. Initially, we specify an arbitrary droplet profile and let surfactants deposit uniformly on the droplet region. More specifically, the initial condition used for the film thickness $h$ and the surfactant concentration $\Gamma$ for $-L \le x \le L$ are given by
\begin{equation}
\begin{aligned}
    h(x,0) &= 0.03 + \max\left\{d_h\left(1-\left({x}/{d_w}\right)^2\right), 0\right\}, \\
    \Gamma(x,0) &= \frac{L}{2d_w}\left(1-\tanh(200(|x|-d_w))\right),
\end{aligned}
\label{eq:ic}
\end{equation}
where $d_h=0.8$ and $d_w=1$ are the initial droplet height and half-width, and $L=1.2$ is the computational domain boundary. 
As the process starts, both $h$ and $\Gamma$ converge to their steady states, where $\Gamma$ reaches a uniform profile due to the diffusion process. 
The simulation is consistent with the analytical solution of equation~\eqref{eq: gamma_eqn}. At the steady state, $\Gamma$ satisfies Laplace's equation with a no-flux boundary condition. Due to the conservation, we have $\Gamma(x,\infty)=1$ for $-L \le x\le L$ at steady states.

\subsection{Autophobing} \label{sec: autophobing}
If pH of liquid is above the isoelectric point, such as water with pH = 6.5 on a silicon dioxide surface, electric charges are induced on the surface and form an electric field that causes autophobing.
We use the steady-state solution from Section \ref{sec: relaxation} as the initial condition for this case. At $t=0_+$, we impose an electric field given in equation~\eqref{eq: pin_electrode} with $\Psi(t)=\Psi_\text{autophobing}=-0.2$, and maintain this electric field induced by surface charges until the system reaches its steady state. 
\begin{figure}[htpb]
    \includegraphics[width=0.42\textwidth]{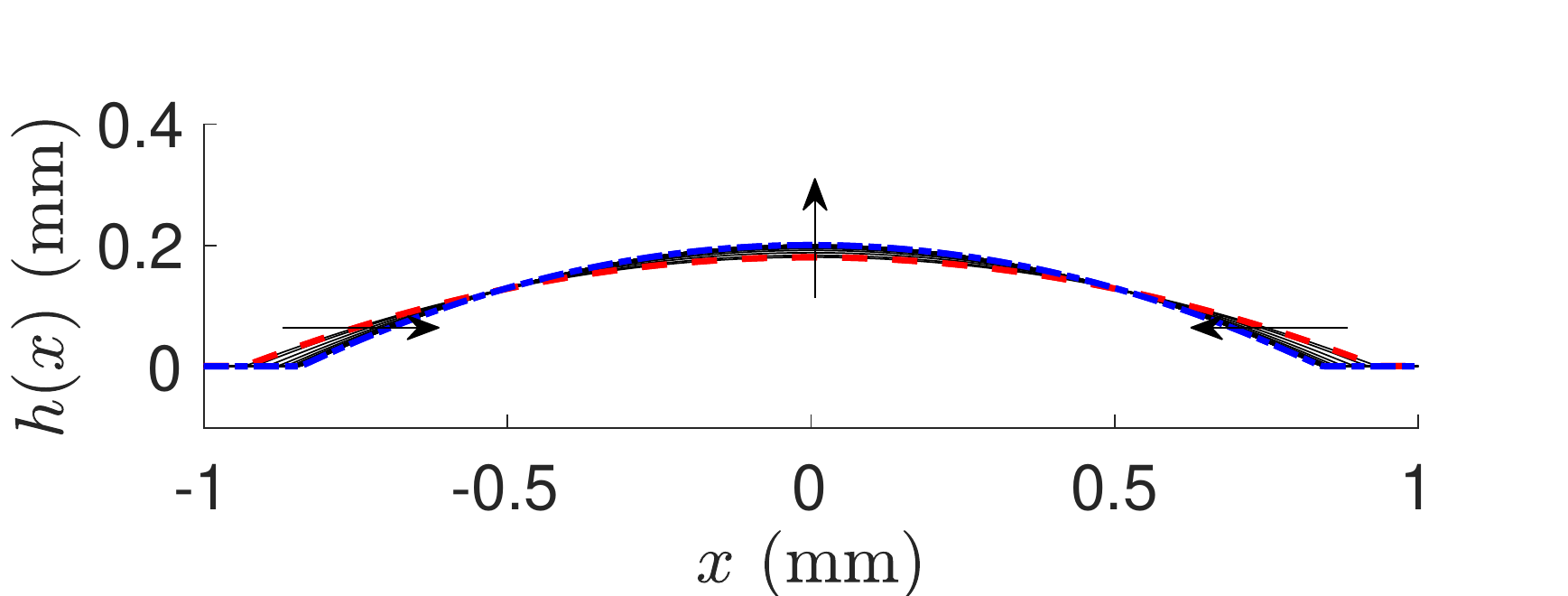}
    \includegraphics[width=0.5\textwidth]{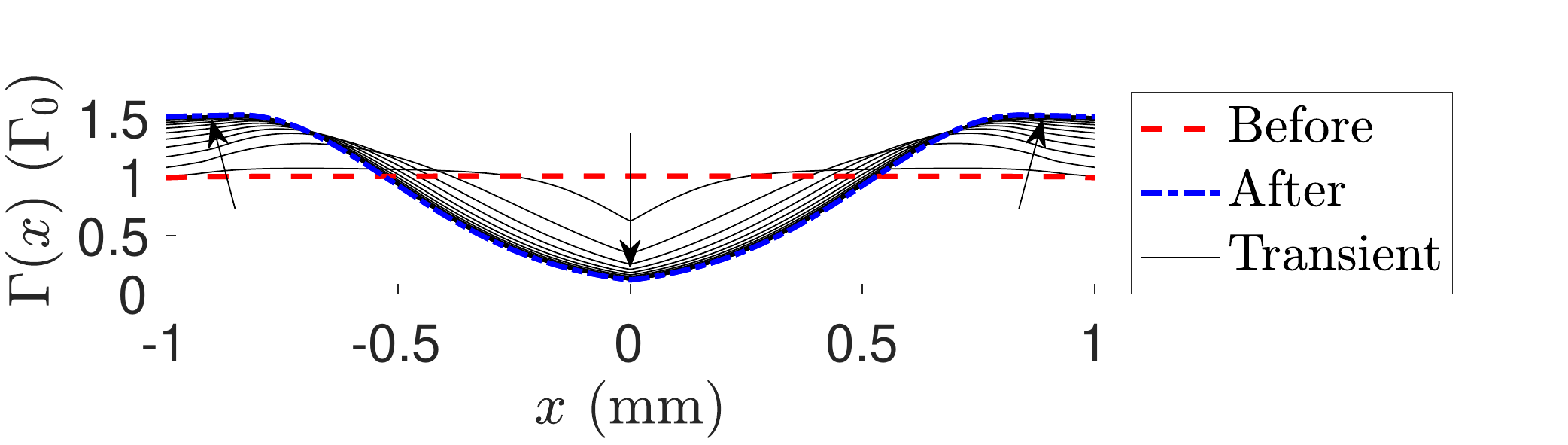}   
    \caption{A comparison between before and after autophobing. The red dashed curves show the initial profiles of the film thickness $h$ and surfactant concentration $\Gamma$ for the autophobing simulations at $t=0$s, which are the same as the blue dashed-dotted curves in Fig. \ref{fig: relaxation}. The blue dashed-dotted curves show the steady-state profiles of $h$ and $\Gamma$ at time $t=5$s. The arrows show the trend of the evolution as time increases. The contact angles are $20.1^{\circ}$ and $24.5^{\circ}$ for before and after autophobing, while the experimental data shows the contact angles are $18.5^{\circ}$ and $25.7^{\circ}$ for before and after autophobing.}
    \label{fig: auto}
\end{figure}

We show the transient dynamics of film thickness $h$ and surfactant concentration $\Gamma$ in Fig.~\ref{fig: auto}. After we apply a negative potential on the substrate, surfactants in the center region move towards the triple point immediately and accumulate in the region. Later on, the accumulated surfactants diffuse out on the precursor layer region and settle to a steady state.
\subsection{Electro-dewetting and electro-rewetting}
This case is motivated by the electro-dewetting and electro-rewetting measurements reported in \cite{li2019ionic}. 
We start the simulations with the steady-state solution from Section \ref{sec: autophobing}, where the droplet has reached its steady state after autophobing. In the first half of the experiment, we impose a negative substrate potential, which induces an electro-dewetting phenomenon and makes the droplet bead up, until the droplet settles to the electro-dewetted steady state. In the second half of the experiment, we reverse the electric field by imposing a positive potential on the substrate, which induces an electro-rewetting phenomenon and makes the droplet spread again until the droplet settles to the electro-rewetted steady state. 
More specifically, the electric potential is imposed as in equation~\eqref{eq: pin_electrode} with
\begin{equation}
    \Psi(t) = \left\{
    \begin{array}{cl}
      -1+\Psi_\text{autophobing},   & \quad  \text{for electro-dewetting},\\
      1+\Psi_\text{autophobing},   & \quad \text{for electro-rewetting},
    \end{array}
    \right.
\end{equation}
where $\Psi_\text{autophobing}=-0.2$ is the electric potential induced by surface charges and maintains along the whole process. 

\begin{figure}[htpb]
    \textbf{(a) Transient dynamics of electro-dewetting \qquad\qquad (b) Transient dynamics of electro-rewetting} \qquad \vspace{0.05em}\\
    \includegraphics[width=0.43\textwidth]{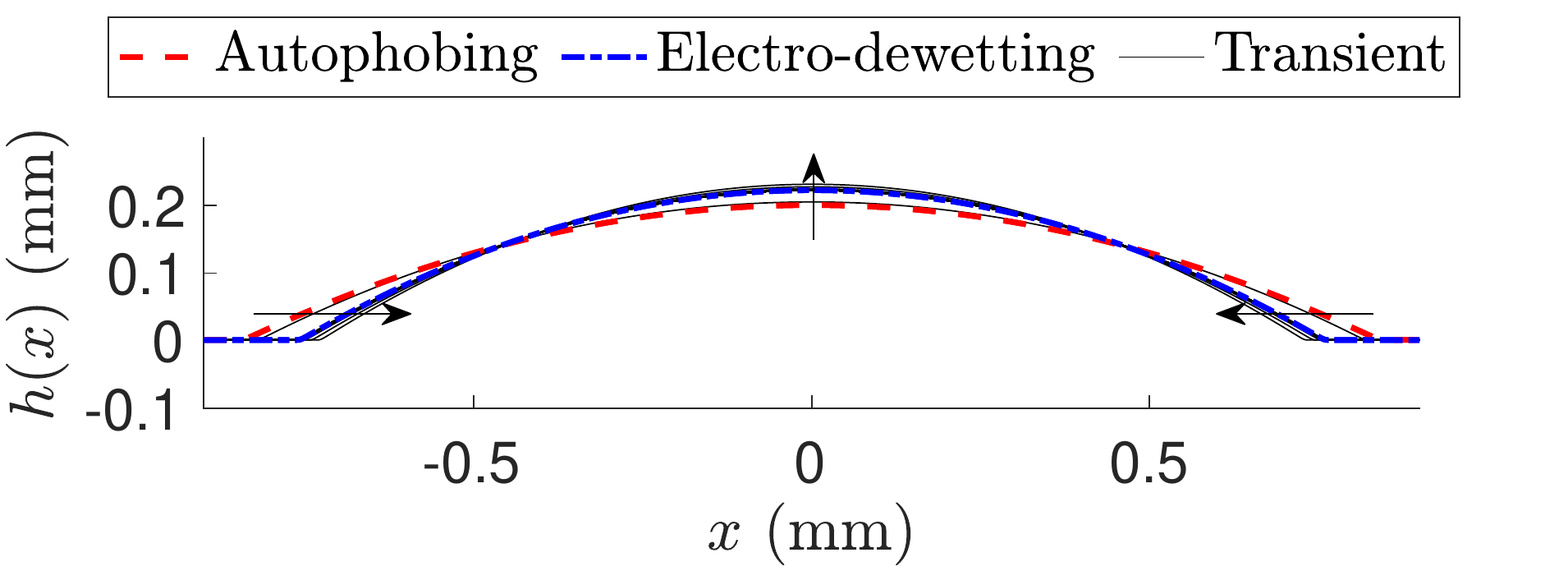} 
    \includegraphics[width=0.52\textwidth]{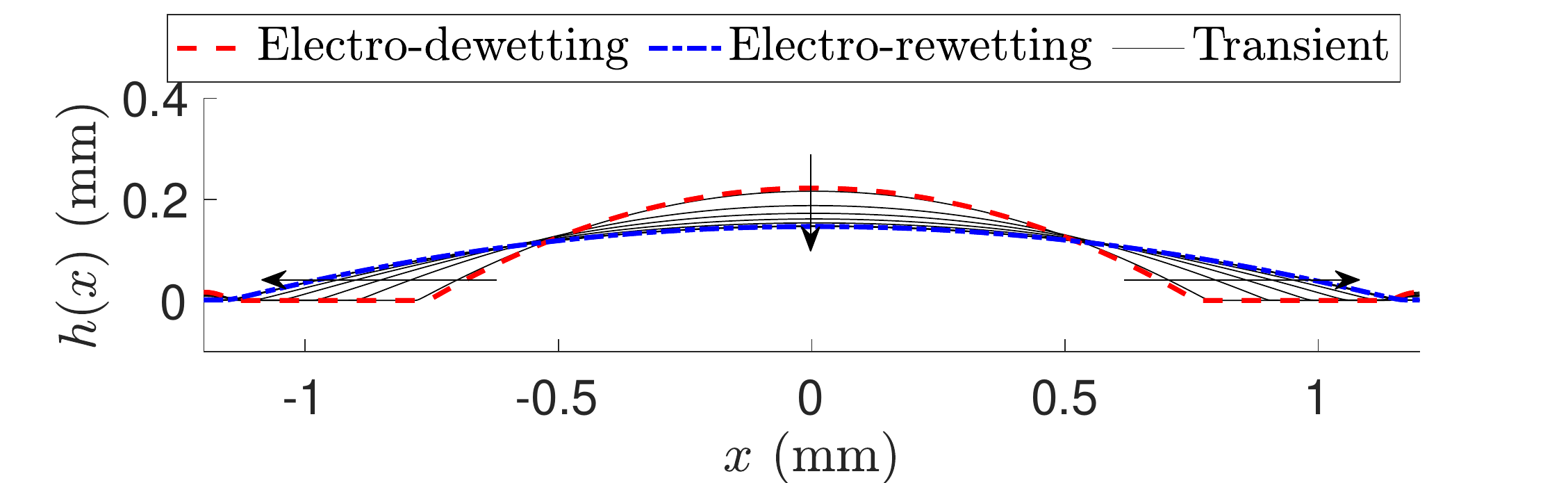}\vspace{-0.1em}\\    
    \includegraphics[width=0.43\textwidth]{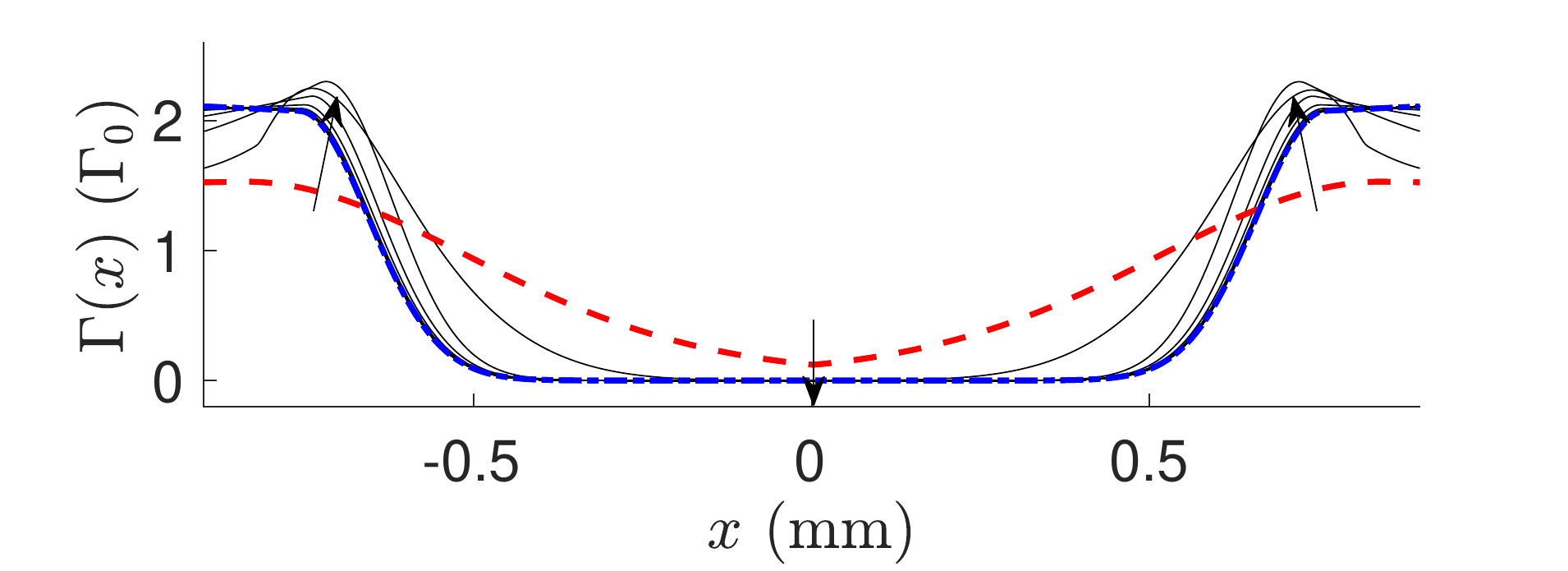} 
    \includegraphics[width=0.52\textwidth]{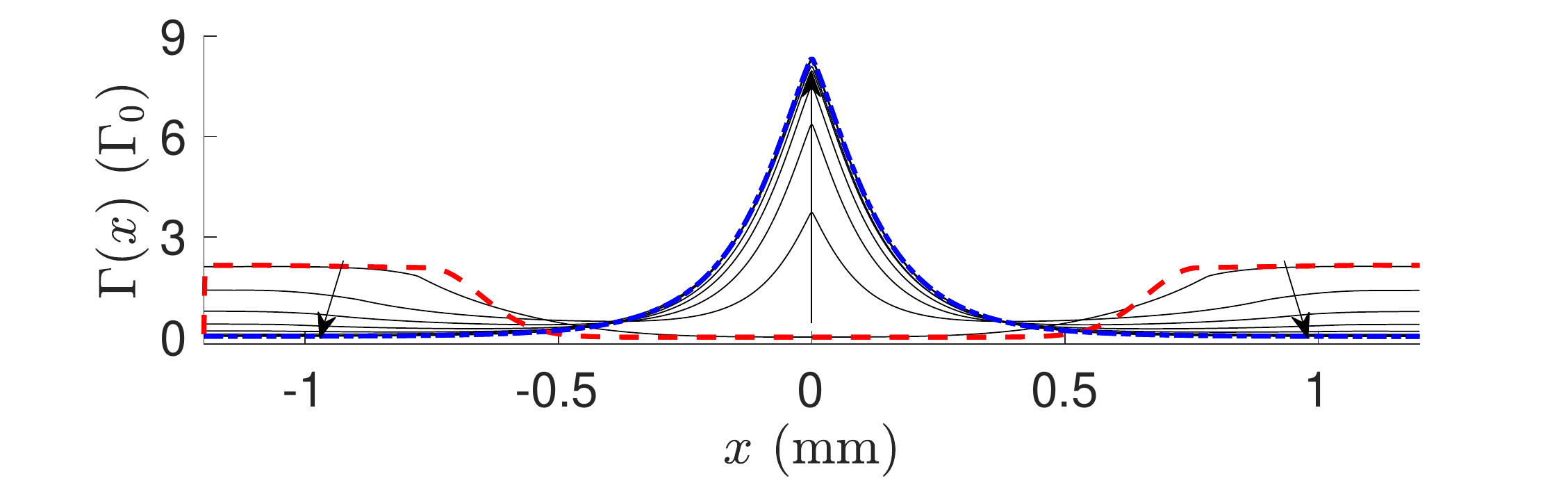}  \\
    \caption{The dynamics of a droplet arriving to its steady state with external electric fields to mimic (a) electro-dewetting and (b) electro-rewetting. The surfactant concentration unit is $\Gamma_0=\text{4.56e-2}~\text{mmol}/\text{m}^2$. 
    For all figures, the red dashed curves show the profiles of $h$ and $\Gamma$ as we start electro-dewetting (or electro-rewetting) simulations (\ie, $t=0$s), and the blue dashed-dotted curves show the steady-state profiles at time $t=5$s for the electro-dewetting (or electro-rewetting) potential. The arrows show the trend of dynamic evolution as time increases.}
    \label{fig: dewet_rewet}
\end{figure}

We present the dynamics of two processes in Fig.~\ref{fig: dewet_rewet}. 
For the electro-dewetting, the cationic surfactant molecules on the droplet are attracted to the triple point region by a negative substrate potential and then the accumulated surfactants gradually diffuse away to a stabilized steady state as the droplet retracts. For the electro-rewetting, which starts with the electro-dewetted steady-state profiles of $h$ and $\Gamma$ as the initial condition, the surfactants are attracted to the top center of the droplet by a positive substrate potential, and then the surrounding surfactants gradually diffuse in toward the triple point region to a stabilized steady state as the droplet spreads. For both electro-dewetting and electro-rewetting, we observe that surfactants on the droplet move immediately after applying the potential since the electric field is strong within the droplet region, while a stabilized steady state is reached much slower by diffusion outside the droplet where the electric field is negligible.

\begin{figure}[htpb]
    \includegraphics[width= 0.85\textwidth]{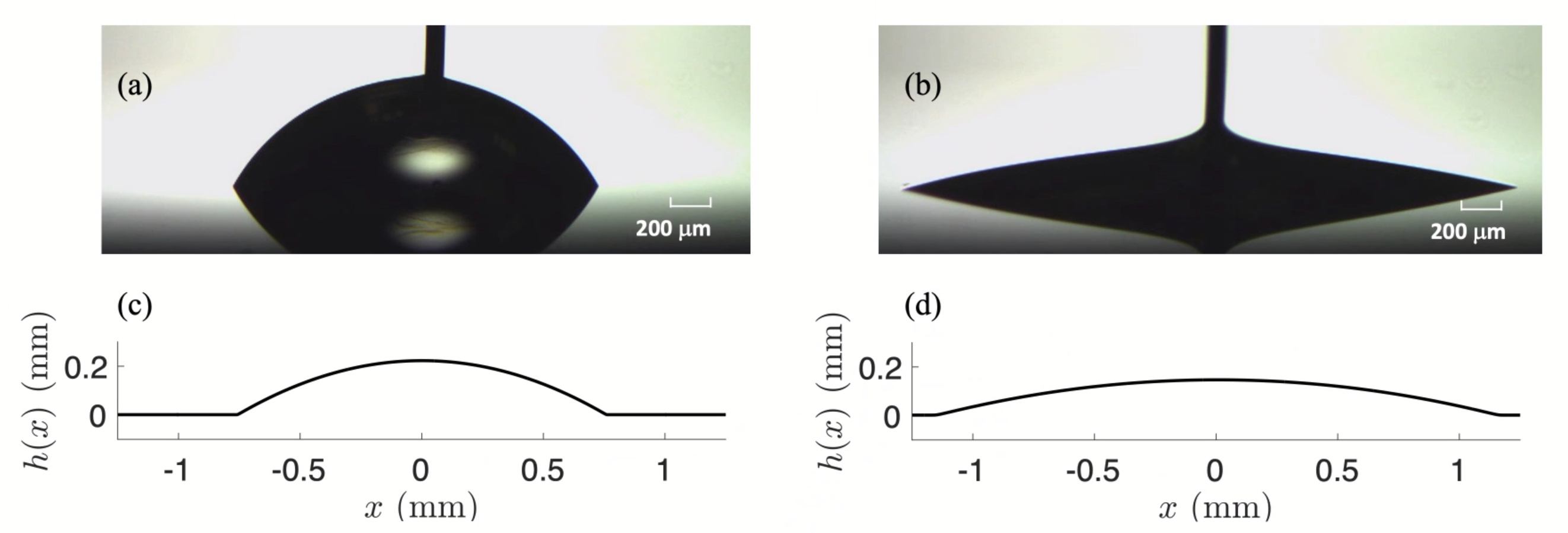}
    \caption{A comparison between experimental and numerical results of droplet shapes for the electro-dewetting (a)(c) and electro-rewetting (b)(d).
    The black solid curves in (c)(d) are the same as the blue dashed-dotted curves in Fig. \ref{fig: dewet_rewet} top panels.
    }
    \label{fig: steady-state}
\end{figure}

We compare the steady-state droplet shapes from the experiment with numerical simulations in Fig.~\ref{fig: steady-state}. The simulation results show that the contact angles are $32.6^{\circ}$ and $13.2^{\circ}$ for electro-dewetting and electro-rewetting cases at steady states, while the experiment measurements are $59.2^{\circ}$ and $14.5^{\circ}$, respectively. 
We note that the electro-dewetting contact angle from numerical simulations is noticeably smaller than the experimental measurement. 
In reality, DTAB (used for the experiment) is a soluble surfactant with its bulk and interface concentrations always balanced. 
When the interface surfactants accumulate near the triple point regions by an electro-dewetting potential, bulk surfactants supplement the interface surfactants, enhancing the aggregation of surfactants on the triple point region. Since the current model does not account for the surfactants in the bulk and omits the supplemental surfactants from the bulk to the contact line region under voltage, thus the simulated contact angle is noticeably smaller than the reality. 

In addition, when the contact angle is around $59.2^{\circ}$, the thin film assumption breaks down and the current linear curvature model is no longer reliable in terms of approximating the thin-film behaviors. One possible approach to improve the comparison with experimental observation is to keep the exact curvature term in the model. Prior works \cite{snoeijer2006free,lopes2018multiple} have shown that this can sometimes improve accuracy. In the present work, we use the linearized curvature term that is more consistent with the lubrication approximation in the asymptotic sense as we neglect the small terms of the same order in the model.
%


\subsection{Droplet shifting}
\begin{figure}[htpb]
    \includegraphics[width=0.85\textwidth]{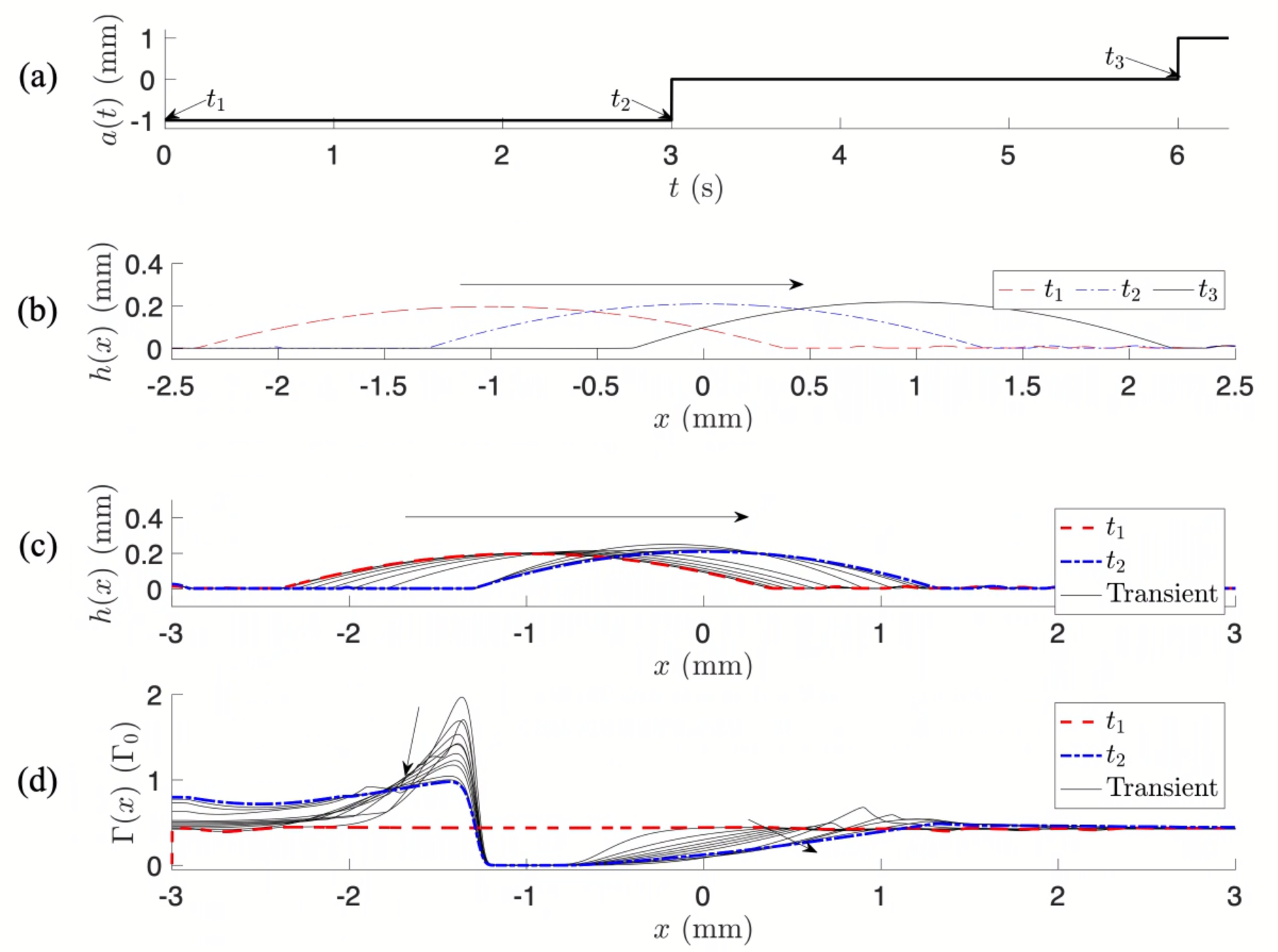} \vspace{0em}
    \caption{A transient process of directional droplet movement with nonhomogeneous electric potential. The arrows in (b)-(d) show the trend of dynamics as time increases.}
    \label{fig: comparison4}
\end{figure}

Imposing a nonhomogeneous potential $\Psi_0(x,t)$ can lead to directional  droplet transportation, which is a fundamental operation in droplet manipulation \cite{cho2003creating}. For this case, we only impose substrate potential which takes the form of a shifted sign function as provided in equations~\eqref{eq: case3_psi0} and \eqref{eq: case2}. 
We prepare the system by placing a droplet centered at $x=-1$, whose shape takes a similar form as in equation~\eqref{eq:ic}, with $d_w=1.3$ and $d_h=0.8$. We let it rest to its steady state without the external potential first, and then turn on the substrate electrode and impose a nonhomogeneous shifted sign function, whose center $x=a(t)$ satisfies
\begin{equation}
    a(t) = \begin{cases}
    & -1, \quad \mbox{for} \quad  0\le t < 3,\\
    &0, \quad \mbox{for} \quad 3 \le t < 6, \\
    &1, \quad \mbox{for} \quad t \ge 6. \\
    \end{cases}
\label{eq:time-varying-Psi}
\end{equation}
We show the center position $a(t)$ in Fig.~\ref{fig: comparison4}(a) and report the dynamics of droplet profiles at three timestamps in Fig.~\ref{fig: comparison4}(b), which corresponds to the steady-state profiles when the nonhomogeneous potential is centered at different locations. 
Fig.~\ref{fig: comparison4} (c-d) show the transient dynamics of film thickness $h$ and surfactant concentration $\Gamma$ between time $t_1$ and $t_2$, which is the first period that we complete the directional move. Initially, the surfactants exhibit a uniform stationary profile without the presence of electric force. After a nonhomogeneous potential is turned on, surfactants quickly move towards the left by the electric force and induce a left contact angle increase, which leads to a dewetting effect; on the other hand, the right contact angle decreases due to surfactants movement and leads to a wetting effect. The piecewise electric potential breaks the symmetry in both the film thickness and surfactant profiles and drives the droplet to move towards right, which has a higher electric potential.


\section{Conclusions and future work}\label{sec: conclusions}
In this work, we propose a microfluidics-based lubrication model to describe the dynamics of a droplet on substrate, manipulated by a direct current (DC) electric field. The model takes into account surface tension, Marangoni effects, electric force, and intermolecular force and explains droplet actuation from an electro-hydrodynamics perspective. This work first considers the surfactant transport under the induction of an electric field and investigates the underlying relation between film thickness and surfactant concentration, which also provides insights into complex phenomena involving interactions between electric fields and fluid mechanics. 

The model starts from a free boundary problem of the Navier-Stokes equation for an incompressible thin liquid film and incorporates the electric effect represented by an approximated electric potential. Surfactants, serving as the main agent to affect the surface tension, are described by one transport equation subject to electric forces. We use a surfactant-dependent disjoining pressure to incorporate the intermolecular forces between surfactants on the solid substrate and the thin liquid film.
Using the lubrication theory, we propose two coupled nonlinear PDEs for the film thickness $h$ and surfactant concentration $\Gamma$ that describe the overall dynamics.
To simplify calculations, we approximate the electric field on the water-air interface and showcase two configurations typically used in droplet actuation experiments via electro-dewetting.
The numerical simulations match the experimental results reasonably well. When a dewetting potential is applied, we observe the surfactants leaving the droplet region and entering the adsorption layer. When a rewetting voltage is applied, we observe the surfactants leaving the adsorption layer and entering the droplet region. We also apply a piecewise constant substrate potential to reproduce directional movement by inducing asymmetric contact angle changes of a droplet. We simulate autophobing, which is typically coupled with electro-dewetting, by modeling the surface charges with a virtual electric potential.

In the current work, assumptions have been made for model simplification and computational convenience, including surfactant insolubility, a two-dimensional droplet assumption, no influence of charged surfactants on the electric field, and parabolic approximation of the spherical droplet shape.
As for future work, we could consider lifting some of these assumptions and investigate another microfluidic operation -- droplet splitting -- which occurs in a three-dimensional space.  

\section*{Acknowledgements}
This work has been partially supported by the National Science Foundation (NSF) (1711708 and 1720499) (C.-J. Kim and Q. Wang) and Volgenau Endowed Chair in Engineering at University of California Los Angeles (C.-J. Kim). This work has also been partially supported by Simons Foundation Math+X investigator award number 510776 (A. L. Bertozzi, W. Chu, H. Ji, and Q. Wang) and Betsy Wood Knapp Chair for Innovation and Creativity at University of California Los Angeles (A. L. Bertozzi).

\section*{Appendix}
\label{sec:appendix}
In Table \ref{tab: nomenclature}, we show the relevant nomenclature and the corresponding value ranges for the electro-dewetting experiments. 
\begin{table}[ht]
\captionsetup{justification=centering}
\caption{\label{tab: nomenclature}Nomenclature and their sample values.}
    \begin{tabular}{lll}
      Definition   &  Symbol \quad & Typical values \\
      Length scale in $x$-direction (mm)  & $L$ & $ 1\sim 3 $\\
      Length scale in $y$-direction (mm)  & $H$ & $ 0.2\sim 0.7 $ \\
      Time scale (s) & $T$ & $1 \sim 3$ \\
      Surfactant concentration on interface (mmol/$\text{m}^{2}$) & $\Gamma_0$ & $5\times 10^{-2}$\\
      Scale of external voltage (V) & $\Psi_0$ & $1\sim 3$\\
      Air-liquid interface diffusion coefficient ($\text{m}^2$/s) \quad & $D_s$ & $10^{-9}$
    \end{tabular}
\end{table}

\bibliographystyle{unsrt}  
\bibliography{bibfile}

\end{document}